\documentclass{article}
\usepackage{epsf,graphicx}
\usepackage{url}

\newlength{\refindent}
\newlength{\parskiplen}
\begin{document}


\newenvironment{references}{\clearpage
			    \section*{\large \bf REFERENCES}
			    \parindent=0mm \everypar{\hangindent=3pc
			    \hangafter=1}}{\parindent=\refindent \clearpage}
\newenvironment{figcaps}{\clearpage
			 \section*{\large  \bf FIGURE CAPTIONS}}{}
\newcommand{\fig}[2]{\parbox[t]{2.0cm}{Figure #1:} \
		   \parbox[t]{13.5cm}{#2}\\[\baselinestretch\parskiplen]}


\begin{titlepage}
\begin{center}
\vspace*{0.5cm}
{\huge Optical Spectra of Ultracool Dwarfs}\\[0.6cm]
{\huge with the Southern African Large Telescope.}\\[3.0cm]

{\large C. Koen$^1$, B. Miszalski$^{2,3}$, P. V\"ais\"anen$^{2,3}$
and T. Koen$^3$}\\[1cm]
\normalsize
{\em 1 Department of Statistics, University of the Western Cape,
Private Bag X17, Bellville, 7535 Cape, South Africa}\\
ckoen@uwc.ac.za \\[0.6cm]
{\em 2 South African Astronomical Observatory, PO Box 9, Observatory, 7935 Cape,
South Africa}\\[0.6cm]
{\em 3 Southern African Large Telescope, PO Box 9, Observatory, 7935 Cape,
South Africa}\\[2cm]

\end{center}

\begin{quotation}\noindent{\bf ABSTRACT.}
New spectra of 81 ultracool dwarfs (spectral types M7 and later) are discussed.
Spectral classifications of 49 objects are available in the literature, while 32
objects are newly classified. The known spectral types were used to test an automated
classification scheme, which relies primarily on template fitting,
supplemented by matching of spectral indices calibrated against the template
spectra. An attempt was made to quantify the uncertainty in the spectral types,
which is generally better than two subclasses. Objects for which spectral types
differ by more than one subclass from the literature classifications
are discussed individually. 
Discrepancies between automated classifications based on 
respectively template fitting 
and spectral index matching, may be useful for flagging objects with unusual spectra.
Aside from the 32 first-time classifications,
alternative classifications are presented for 32 previously classified dwarfs. 
Very large (equivalent width greater than 130 \AA) H$\alpha$ flares are 
reported for the known
ultracool dwarf binary 2MASS~J15200224-4422419;
curiously, the object does not appear to have quiescent emission lines. 
Non-zero equivalent width measurements are listed for a further 29 objects.

\vspace*{1.0cm}
{\bf Key words:} stars: low-mass - stars: flare - stars: brown dwarfs 
\end{quotation}

\end{titlepage}

\section*{\large \bf 1 INTRODUCTION}

This paper presents far red (6300-8800 \AA\ ) spectroscopic observations 
of 81 ultracool
dwarfs (UCDs), i.e. objects (stars and brown dwarfs) with spectral types
of M7 or later. The aims of this ongoing project are to (i) obtain 
classification spectra
of photometrically selected candidate UCDs; (ii) obtain optical spectra
of UCDs for which only NIR spectral classifications have been published; 
(iii) to revisit confirmed UCDs in order to investigate the prevalence of
spectral variability; and (iv) to obtain H$\alpha$ equivalent widths of a
large sample of UCDs.

A consequence of the low temperatures of UCDs is that objects
with these spectral types are much brighter at infrared wavelengths, which
has no doubt contributed to the large number of near infrared (NIR) spectra
which have been published -- the comprehensive websites
\url{http://spider.ipac.caltech.edu/staff/davy/ARCHIVE/index.shtml} and
\url{https://jgagneastro.wordpress.com/list-of-ultracool-dwarfs/}
contain references to hundreds of examples. However, spectral classification for 
all stars and brown dwarfs that could
be reasonably detected optically (which includes late M and L type UCDs), was
historically developed for spectra taken in the optical (e.g. Kirkpatrick et al. 1999).
Added to this 
are the facts that the spectral sequence is not as obvious in the NIR, and
that NIR colours at a given spectral type show considerable scatter -- see Allers
\& Liu (2013) for a discussion. As a result, classification based on
NIR spectra is less certain (e.g. Luhman et al. 2003, Lodieu et al. 2005).
[Note though that although definitive classifications may not
always be obtainable from NIR spectra, these do contain
further information about conditions in UCD atmospheres, since different
physical environments are sampled than by optical spectroscopy 
(Kirkpatrick 2009)].

Modest spectral resolution is sufficient to obtain H$\alpha$ equivalent
widths, and this is exploited in the study reported below. 
A recent discussion of magnetic activity in UCDs can be found in Metodieva
et al. (2015), which is also a source of references to earlier work (see
also section 5 of this paper). The authors provide a plot of the activity
indicator log~$L_{{\rm H}\alpha}/L_{\rm bol}$ against spectral type, 
based on their own observations and values taken from the literature.
There is a clear decrease in mean activity, the later the spectral type.
Since relatively strong magnetic fields have been observed in some 
late M/early L UCDs (e.g. Williams, Cook \& Berger 2014, and references therein), 
the activity decrease may be due to the inability
of largely neutral atmospheres to manipulate the magnetic fields. 
An exception to this general trend is discussed in section 5, in the 
form of an enormous flare associated with a binary UCD consisting of
two early L dwarfs. It is particularly noteworthy that other visits to
this object did not reveal any signs of quiescent H$\alpha$ emission.

The paper presents spectra of 81 objects, with 32 new classifications of
photometric UCD candidates taken from Folkes et al. (2012). Spectral types 
for the remaining 49 UCDs were
available in the literature, and were used to thoroughly test our classification
software. Spectral typing is performed by comparison of SALT spectra to sets
of template spectra, and rely on matching either continuum shapes (template
fitting) or sets of spectral indices.

\section*{\large \bf 2 OBSERVATIONS}

We obtained spectroscopic observations of many candidates with the Robert 
Stobie Spectrograph (RSS; Burgh et al. 2003; Kobulnicky et al. 2003) on the 
queue-scheduled Southern African Large Telescope (SALT; Buckley, Swart 
\& Meiring 2006; O'Donoghue et al. 2006) under programmes 2014-1-RSA-003, 
2014-2-MLT-003 and 2015-1-MLT-003. The observing strategy involved 
providing a large surplus of targets with relatively short exposure 
times of 2$\times$600 s across large parts of the sky such that 
observations could readily be taken during any gaps in the SALT queue. 
This filler program approach had relatively loose constraints on the 
seeing and sky conditions to further maximise the chance that observations 
would be taken. A red RSS configuration was adopted with the PG900 grating 
covering the wavelength range of $\sim$6320--9268 \AA. The slit widths 
varied between 1.00, 1.25 and 1.50 arcsec, depending on the target 
brightness, yielding resolving powers of 2128, 1703 and 1419, respectively, 
at the central wavelength of 7828 \AA. The reciprocal dispersion was 
0.94 \AA\ pixel$^{-1}$.

Table 1 gives the Two Micron All Sky Survey (2MASS) 
point source catalogue designation (Skrutskie et al. 2006),
date and exposure time of the observations used in this paper. 
A neon-argon or xenon arc lamp reference spectrum was taken after 
each sequence of science exposures, as well as five internal flat-field 
lamps exposures. Data reduction was performed 
using \textsc{iraf}\footnote{IRAF is 
distributed by the National Optical Astronomy Observatory, 
which is operated by the Association of Universities for Research in 
Astronomy (AURA) under a cooperative agreement with the National 
Science Foundation.} on the SALT pipeline data products processed 
by \textsc{pysalt} (Crawford et al. 2010). The data were trimmed,
leaving only the central region of each frame, and 
then cleaned of cosmic ray 
events using \textsc{lacosmic} (van Dokkum 2001). The CCD chips were 
interpolated over before flat-fielding the data by a normalised 
average flat-field that had first been divided by a smoothed median 
of itself. Wavelength calibration was performed using the 
standard \textsc{iraf} tasks \textsc{identify}, \textsc{reidentify}, 
\textsc{fitcoords} and \textsc{transform}. Before extracting 
one-dimensional spectra with \textsc{apall}, all acquisition 
images from SALT were carefully inspected, along with images 
from the SuperCOSMOS Sky Survey (Hambly et al. 2001) and 
2MASS (Skrutskie et al. 2006), to ensure the correct target was on the 
slit and to identify its spatial position on the CCD. This step was 
necessary given the challenging acquisition task of acquiring very 
red targets with high proper motions in sometimes very crowded fields. A 
relative flux calibration was applied to all spectra using spectrophotometric 
standards taken during each semester of observations in the standard fashion. 
The wavelength range was then trimmed to $\sim$6320--8800 \AA\ to remove the 
reddest wavelengths affected by second-order contamination introduced by 
the grating. Where possible a combined spectrum for each object was 
created from a simple average of the two or three spectra taken during 
each individual observation. 

H$\alpha$ equivalent widths (EWs) were measured by fitting a Gaussian profile
to the emission line using the \textsc{IRAF} task \textsc{SPLOT}.

\section*{\large \bf 3 SPECTRAL CLASSIFICATION: PROCEDURES}

Before analysis, the spectral regions 6860-6940, 7590-7680, 7300-7365 and
8310-8365 \AA\ were removed; the first two are affected by the telluric
Fraunhofer A and B features, the latter two include the gaps between CCD 
detectors.

\subsection*{\bf 3.1 Template fitting (TF)}

The primary classification tool was the fitting of template spectra: the overall
shapes of the spectra are 
more resistant to the effects of noise, and
less affected by secondary effects such as gravity variations. Templates were
obtained from Bochanski et al. (2007) (M dwarfs) and Schmidt et al. (2014)
(L dwarfs). The wavelength spacing of the L dwarf templates varies from
0.36 \AA\ for early L, to 1.1 \AA\ for late L. This is similar to the 0.94 \AA\
spacing of our spectra, so that template spectra were simply interpolated.
M dwarf template spectra are given at a finer wavelength resolution of 
0.1 \AA\, and these were binned to 0.9 \AA\ for comparison with the SALT spectra.

Assuming that flux calibrations have been made, 
matching SALT and template spectra
consisted solely of determining a scale factor $\beta$. This is 
determined by least squares as
$$\beta=\sum_\lambda F_S(\lambda) F_t(\lambda)/\sum_\lambda F^2_t(\lambda)$$
where the subscripts $S$ and $t$ respectively denote the SALT and template
spectra. The ``figure of merit" is the sum of squares
\begin{equation}
SS_1=\sum_\lambda [F_S(\lambda)-\beta F_t(\lambda)]^2 \; .
\end{equation}

\subsection*{\bf 3.2 Spectral indices matching}

Secondary classification tools are based on spectral indices.
Twenty two indices were taken from Reid, Hawley \& Gizis (1995), 
Kirkpatrick et al. (1999), Hawley et al. (2002) and Cruz et al. (2009)
-- see Table 2. For the first 17 indices, the ratios 
of the fluxes in the two wavelength intervals were calculated. The last four
indices are defined as the mean flux over the first interval, divided by
the mean of the fluxes over the two denominator intervals. The CaH1 index
was calculates as described in Reid et al. (1995), by interpolating the mean flux
between the two denominator intervals. 

The indices were calibrated by 
calculating each of them for all of the 19 template spectra.
The results are in Figs. 1-4, the first of which also shows 
(as red lines) digitisations
of the fits plotted in fig. 3 of Hawley et al. (2002). Inspection of the diagrams
shows that the late L template spectra are not of sufficient quality to calibrate
the K-b index. Furthermore, it is not clear whether the calculated CaH1 index is
accurate enough to be usable -- if it is, it may be of limited utility anyway, since 
the plot suggests that its discriminatory power may be limited. The K-b and 
CaH1 indices were therefore not used for classification.

To elaborate: the CaH1 index involves fluxes in the wavelength interval 6345-6420
\AA, some 250 \AA\ blueward of the wavelength ranges of all other indices
in Table  2. Given the extreme dearth of blue flux emitted by UCDs with
mid and late L spectral types (see e.g. Fig. 6), it is no surprise
that it is difficult to accurately calibrate the CaH1 over the entire
spectral range considered in this paper. At first sight the K-b index
may be expected to be reliable at later spectral types. However, the
broad K~I absorption trough due to the doublet at 7665 and 7699 \AA\ deepens
with deceasing temperature and by mid L the flux level in this part of the 
spectrum is comparable to that on which the CaH1 index is based. In some
of the later type template spectra fluxes in the wavelength ranges of these
two indices are close to zero, or even negative.

Noise sometimes leads to calculated indices being outside the ranges covered by
the calibrations. If the discrepancy is a few percent, then it seems
plausible to associate the index with the closest spectral class. (For example, if
TiO4=0.5, the estimated spectral class is M7). Otherwise, if the margin is too large,
the index is rejected. 

The most straightforward classification based on indices is to search for the 
spectral type
with the best overall agreement between observed and calibrated indices:
\begin{equation}
\min_t [SS_2(t)]=\min_t \sum_j [I_j({\rm observed})-I_j({\rm template} \; t)]^2
\end{equation}
where $I_j$ denotes spectral index $j$, and $t$ again indexes the template spectra. 
An alternative which is more
resistant to outlying index values is
\begin{equation}
\min_t \sum_j |I_j({\rm observed})-I_j({\rm template} \; t)| \; .
\end{equation}
In order to place all indices on equal footing the $I_j$ are standardised such
$0 \le I_j({\rm template}) \le 1$ (i.e. the minimum value for the particular 
template index is subtracted, and the result is divided by the range spanned
by the index).

Approximate standard errors of the spectral indices follow from the delta method as
\pagebreak

\begin{eqnarray}
{\rm S.E.}(I_j) &=& \left [ \frac{{\rm var}(F_n)}{\ell_n \overline{F}_d^2}+
\frac{\overline{F}_n^2 {\rm var}(F_d)}{\ell_d \overline{F}_d^4} \right ]^{1/2} 
\;\;\;\;\; j=1-17\nonumber\\
 &=& 13\left \{ \frac{{\rm var}(F_n)}{\ell_n (7\overline{F}_{d1}+6\overline{F}_{d2})^2}+
\frac{\overline{F}_n^2} {(7\overline{F}_{d1}+6\overline{F}_{d2})^4} 
\left [ \frac{{7^2\rm var}(F_{d1})}{\ell_{d1}}+\frac{6^2{\rm var}(F_{d2})}{\ell_{d2}}\right ]
\right \}^{1/2} \;\;\;\;\; j=18\nonumber\\
 &=& 2\left \{ \frac{{\rm var}(F_n)}{\ell_n (\overline{F}_{d1}+\overline{F}_{d2})^2}+
\frac{\overline{F}_n^2} {(\overline{F}_{d1}+\overline{F}_{d2})^4} 
\left [ \frac{{\rm var}(F_{d1})}{\ell_{d1}}+\frac{{\rm var}(F_{d2})}{\ell_{d2}}\right ]
\right \}^{1/2} \;\;\;\;\; j=19-22 \;.
\end{eqnarray}
Here $\ell_n$ and $\ell_d$ are respectively the number of spectral elements in the
numerator and denominator intervals, while $F_n$ and $F_d$ are the fluxes in the numerator
and denominator intervals. Information about standard errors for the sample of 232
individual SALT spectra is summarised in Fig. 5. Plotted are the fractions of indices
with standard errors smaller than 0.2 (broken line) or smaller than 0.1 (solid line).
Judging by this, indices 2-5 (CaH2, CaH3, TiO2, TiO-a), 7 (TiO5), 9 (VO7434), 11 (VO7912), 
15 (TiO844) and  16 (CrH-a) are particularly useful for
the sample of UCDs dominated by late M to early L objects. Note that 
with the exception of index 12 (Na8190) this list includes 
the four indices proposed by Hawley et al. (2002) which are within the SALT spectral range.
Cruz et al. (2009) fig. 4 shows 8 alkali-metal indices useful for discriminating
gravity effects: of these, Cs-b is outside the spectral range of SALT; the 
denominator of
the K-a index is within the excluded Fraunhofer A wavelength range; K-b is poorly calibrated
by the template spectra; and all the remaining indices (Rb-a, Rb-b, Na-a, Na-b, Cs-a)
are prone to high noise levels. On the other hand, three of the four molecular gravity
indicators in Cruz et al.(2009) fig. 5 (VO-a$\equiv$TIO7434, TiO-b$\equiv$TiO8440 and CrH-a) 
are in the list of useful indices given above.     
In what follows indices with estimated standard errors in excess of 0.2
were not taken into account. 

\subsection*{\bf 3.3 Median index fitting (MIF)}

An alternative method using spectral indices, based on a median spectral
type, was also used. The approach is designed to limit the impact
of spurious classifications supported by few indices. The methodology is 
as follows:
\begin{itemize}
\item[(i)]
Note that for a given index, possible spectral classes are 
essentially read from the 
relevant plot in Figs. 1-4, i.e. spectral types are real, rather than integer.
\item[(ii)]
Since spectral type is 
typically a multi-valued function of spectral index, several
spectral subclasses are usually associated with each index. (For example,
spectral types L1.0, L2.8, L5.6 and L7.3 all have TiO4 indices near unity).
This clearly leads to ambiguity. At the outset all spectral types implied by 
{\it each} of the spectral indices are collected.
\item[(iii)]
Possible classifications are 
narrowed down by rejecting from the list determined in (ii)
all types more than three subclasses from the  TF classification.
\item[(iv)] 
The median spectral subclass of the remaining candidate types 
is then selected as the best index-based classification.
\end{itemize}
 
In all
but 17 cases types derived in this fashion are within 1.5 subclasses of the template
value; this level of agreement is better than that obtainable with the
index-based methods defined by (2) or (3).

\subsection*{\bf 3.4 Classification confidence}

It is, of course, desirable to have some indication of the reliability of 
classifications. It is tempting to use some sort of $\chi^2$ procedure to compare
the different values of $SS_1$ or $SS_2$ [in Eqns. (1) or (2)]
over different spectral class fits. Formally this could 
be done using a procedure such as Bartlett's $\chi^2$ test (Bartlett 1937). However,
it is well known (e.g. Conover, Johnson \& Johnson 1981) that this procedure is 
sensitive to deviations from Gaussianity -- i.e. a significant value of the statistic 
may simply be due to the fact that fit residuals are non-Gaussian. A simple 
alternative is to
use a permutation procedure. For template fitting, based on (1), it would work
as follows:
\begin{itemize}
\item[(i)]
Choose a suitable test statistic, e.g. 
\begin{equation}
U_t=SS_1(t)/\min_t SS_1(t)\equiv SS_1(t)/SS_1(t_0) 
\end{equation}
i.e. $t_0$ is the index of the best-fitting template, and $t$ the index of a 
candidate alternative.
\item[(ii)]
Define the two sets of residuals 
\begin{eqnarray}
e_0(\lambda)&=&F_S(\lambda)-\beta_0 F_0(\lambda)\nonumber\\
e_t(\lambda)&=&F_S(\lambda)-\beta_t F_t(\lambda)\nonumber
\end{eqnarray}
where $F_0$ is the best-fitting template, $F_t$ the candidate alternative, and the 
wavelength $\lambda$ takes on all
the observed values. In terms of the residuals 
\begin{equation}
U_t=\sum_\lambda e_t(\lambda)/\sum_\lambda e_0(\lambda)
\end{equation}
\item[(iii)]
Pool the values of $e_0$ and $e_t$ and randomly divide them into two new
sets of values $e_0^\prime$ and $e_t^\prime$. 
\item[(iv)]
Calculate the statistic $U_t^\prime$ as
defined in (6), using $e_0^\prime$ and $e_t^\prime$. 
\item[(v)]
Repeat steps (iii) and (iv) many (preferably a few thousand) times, saving
the new value of $U_t^\prime$ for each replication.
\item[(vi)]
The significance level of $U_t$ calculated in (i) is then established by
noting its rank in the collection of $U_t^\prime$.
\end{itemize}

The method sketched above has the virtue of being completely distribution-free,
but the drawback of being time consuming if many observed spectra are processed.
A compromise is therefore made: the non-parametric variance comparison test
of Fligner \& Killeen (1976) is used instead. Limited experimentation suggests
that the permutation test, the Fligner \& Killeen (1976) procedure, and the 
more widely used Brown-Forsythe test (Brown \& Forsythe 1974) all give
comparable results (see also Conover et al. 1981). The acronym ``FK" will
be used below to refer to the Fligner \& Killeen
(1976) test.

The question remains as to which significance level to conduct the test at. 
Testing is unconventional,
since the aim is to ascertain which templates (or sets of spectral indices)
provide viable alternatives to the best-fitting one. The value adopted 
was $p=0.1$, i.e. a 10\% chance of obtaining such a large variance for the 
alternative classifications, as compared to the variance
associated with the best-fitting template or spectral index set. (Quite 
similar results are obtained with $p=0.2$). For convenience,
the term ``confidence set" will be used for the collection of spectral classes
which are statistically equivalent (as defined above) to the best-fitting one.

\subsection*{\bf 3.5 Comparison of the three classification methods}

It comes as no surprise that the confidence sets
are smaller for template fitting than for fitting based
on spectral indices: the number of points being fit is two orders of magnitude
greater in the former case. Although calculation of the spectral indices involves
taking averages, the wavelength intervals are relatively narrow, and taking ratios
inflates uncertainties. Only 20 of the template fits to the 232 
individual SALT spectra have 
confidence sets larger than 3 subclasses, whereas 102 of the spectral 
index-based classifications have confidence sets greater than 5 subclasses.

To summarise: the most dependable classification tool is TF, followed
by the restricted median of the collection of spectral types (MIF), as 
derived from the spectral indices. There are very few cases where 
results from TF and MIF are incompatible, and in those cases the disagreement is
marginal, as demonstrated by MIF spectral types being in the confidence sets of 
the TF
classifications. Unrestricted index matching, as described in subsection 3.3,
is very inefficient as compared to the other two methods. The reason, alluded 
to in subsection 3.3, is that spectral type is not a unique function of 
spectral index for {\it any} of the indices in Figs. 1-4. Restricting the
part of the multi-valued function to be considered ameliorates this problem.
Furthermore, the median of the spectral types from all the indices is
more representative of the information content of the set of indices than the 
mean spectral type.  
   
A few representative examples of templates fitted to SALT spectra 
can be seen in Fig. 6.

\section*{\large \bf 4 SPECTRAL CLASSIFICATION: RESULTS}

There are 73 averaged SALT spectra of 49 objects with spectral classifications
provided by the Simbad database; these were used to test performance of our 
classification software. A graphical comparison between the literature spectral
types and those obtained from SALT TF can be seen in Fig. 7. (The SALT spectral
types have been randomly offset by small amounts, to avoid many points being 
coincident). Circles denote optical literature spectra, while squares indicate
infrared spectra. Interestingly, there does not seem to be a systematic difference 
between our classifications and Simbad types based on infrared spectra, although the
scatter around the equal-types line is larger for the infrared spectra.

Agreement between TF classifications and those listed in Simbad is generally
good, with 88\% of types differing by less than 2 spectral subclasses. The figure for
MIF classifications is similar -- 78\%. SALT TF types differ by two or more 
spectral classes from Simbad classifications for 8 UCDs -- see Table 3:\\
{\bf 2M~0128-5545} 
\noindent
Kendall et al. (2007) assigned a NIR spectral class of L1, but remarked that some
spectral indices suggested a type about 2 subclasses later. 
Mart\'{\i}n et al. (2010) also classified the UCD as L1. The template fit to the
SALT spectrum suggests an unambiguous L3 type, and this is confirmed by the
spectral indices (20 of which were usable).\\
{\bf 2M~0223-5815} 
\noindent
Cruz et al.(2009) classify this object as L0$\gamma$, meaning 
``very low gravity L0". 
The $p$-value of the FK test statistic is 0.59 for the L1 spectral
fit. It may be concluded that the template fit to the SALT
spectrum is not unique, and does not markedly contradict the L0 classification
in the literature.  \\
{\bf  2M~0230-0953} 
\noindent
Widely different classifications of 2M~0230-0953 have been published
by respectively  Mart\'{\i}n et al. (2010) (optical spectrum, L0) and Marocco et al. (2013) 
(NIR spectrum, L6). The three individual SALT spectra give classifications consistently 
in the range
L1-L3. We adopt a mean classification of L2.\\
{\bf 2M~1523-2347 and 2M~1548-1636}
\noindent
Classifications of L2.5 and L2 were assigned to 2M~1523-2347 and 2M~1548-1636 
respectively by Kendall et al. (2007), on the basis of NIR spectra. All SALT spectra
give consistent results
of L0 for both objects. Given the remarks on NIR typing in the Introduction,
the latter classification is recommended.\\
{\bf 2M~1655-0823}
\noindent
This is a very well-studied object (current Simbad citation count is 333), 
better known by the names GJ~644C, LHS~429 and VB~8. Use of the spectrum as an
M7 standard appears to date back to Boeshaar \& Tyson (1985). There have been some 
remarks about abnormalities in the spectrum (Dahn, Liebert \& Harrington 1986).
SALT spectra of 2M~1655-0823 were obtained at two epochs; M7 and M9 templates
fitted to these are plotted in Fig. 8. It is immediately apparent that
whereas the M7 template fits the strong features in the range 8000-8500 \AA\ quite
well, the M9 template fits the continuum blueward of about 7800 \AA\ much better.
It is noteworthy that the best classification obtained from fitting spectral indices
[Eqn. (2)] is M7, for both spectra. It is cautiously concluded that whereas the spectral
indices are those of an M7 dwarf, the shape of the continuum is unusual, and
not fitted perfectly by any one standard form.\\
{\bf 2M~2252-1730} 
\noindent
The T0 classification listed in Simbad is a revision by Schneider et al. (2014),
based on a NIR spectral
classification,
of the spectral type assigned by Reid et al (2006). The
latter authors found the object to be a binary with L6$\pm1$ and T2/3 
components. The SALT optical spectral type is comparable with the
Reid et al. (2006) NIR
classification of the brighter component.\\
{\bf 2M~2255-5713} 
\noindent
The object is a binary composed of L6 and L8-T1 components (Reid et al. 2008a).
Interestingly, both SALT spectra show an even greater excess at shorter
wavelengths than that visible in Reid et al. (2008)'s fig. 2. 
This clearly cannot be ascribed to the fainter companion, but may be due to
contamination by a faint background star.\\
 
It should be clear from the discussion that there is little reason to doubt the
automated classification of the SALT spectra.
New proposed classifications for some of the objects are given in the 
penultimate column
of Table 3.

It is perhaps worthwhile to also consider those spectra for which the difference between
the TF classification and that extracted from the Simbad database is 1.5 subclasses.
Information about
the spectra, corresponding to 9 different UCDs, appears in Table 4:\\
{\bf 2M~0651-1446 and 2M~1308-4925}
\noindent
The existing classifications of these objects are due to Folkes et al. (2012),
based on NIR spectra. Template fitting of 2M~0651-1446 select a spectral type of M9 
with M8 a very close second.
MIF spectral types are also around M8. A classification of M8.5 seems appropriate. 
As far as 2M~1308-4925 is concerned, the M7 template fits are considerably better
than any competitor, while the indices are consistent with types in the
range M7.3-7.4. A classification of M7 is adopted.\\
{\bf 2M~0814-4020}
\noindent
There is good agreement between the TF and MIF types derived from the SALT
spectra, giving a classification in the range M8-M9. The only published
classification, by G\'{a}lvez-Ortiz et al. (2014), is M7-M8. The proposed
classification is M8.5.\\
{\bf 2M~0847-1532}
\noindent
On discovery the object was classified L2 (Cruz et al. 2003), and was listed as an
L2 spectral standard by Reid et al.(2008b). The Simbad spectral classification
of L1.5 is from a NIR classification by Schneider et al. (2014). As indicated
in Table 4, L2-L3 templates fit the data almost equally well, showing compatibility
with the previous classifications.\\ 
{\bf 2M~1454-6604}
\noindent
The Simbad classification of L3.5 is taken from Phan-Bao et al.
(2008). The most likely classification of the two individual SALT spectra
is L5, with L6 distant second choices. A spectral class of L5 is also
supported by the MIF.\\
{\bf 2M~1520-4422} 
\noindent
This UCD has attracted a fair amount of attention. It is a spatially resolvable binary 
($\sim 1.2$ arcsec separation). Spectral classifications of L2+L4 (Kendall et al. 2007)
and L1.5+L4.5 (Burgasser et al. 2007) have been made, based on NIR spectra. 
Classification by spectral indices lie in a narrow range of L2.2-L2.3.\\
{\bf 2M~1845-6357}
\noindent
This UCD, better known as SCR~J1845-6357A, was classified as an M8.5 dwarf by
Henry et al. (2004), based on an 8.7 \AA\ resolution optical spectrum.
The two $S/N=100$ SALT spectra summarised in Table 3 suggest an alternative 
classification:
although the FK statistic rejects alternative fits, the Brown-Forsythe statistic
indicates that the M9 template fits are not much worse than L0. Bearing in 
mind also the MIFs, an M9.5 classification is proposed.\\
{\bf 2M~2224-0158}
\noindent
The Simbad spectral type of L4.5 is from Kirkpatrick et al. (2000). Fits
of the L4 and L5 templates are comparable to the best fit (L6), and the 
MIF type is L6.2. There is enough uncertainty to accommodate the published
spectral type. \\
{\bf 2M~2308-2722}
\noindent
Templates in the range L1-L3 
fit almost equally well, i.e. the classification formally accords
with the literature spectral type.\\

It may be concluded that in general the classifications in this paper are
within one subclass of previous classifications. Greater differences can largely
be ascribed to uncertainties (as reflected in the confidence
sets) or sometimes large differences between optical and NIR classifications.
Table 5 lists classifications of the remaining SALT spectra of UCDs
with published spectral types: in a few cases slightly different
classifications are proposed. 

First spectral classifications are provided in Table 6 for 32 UCDs.
These were taken from the catalogue of Folkes
et al. (2012), who selected UCD candidates near the galactic plane.
The authors used a combination of
2MASS NIR photometry, optical photometry extracted from photographic surveys,
and proper motion information, to construct their catalogue.

In the majority of cases two SALT spectra obtained in succession have been
averaged. In a few cases spectra were obtained at more than one epoch
(see the dates in column 2).
There is only one object in the Table for which the TF and MIF types 
differ by more than 1.3 classes, namely
2M~0819-4706. This is one of the seven UCDs for which the confidence set consists of three
spectral types, i.e. the fit is more uncertain than most. The spectral index
classification agrees with the latest of the three types in the TF confidence set.
The mean number of indices used for classification (i.e. with estimated standard
errors smaller than 0.2) is 16, and only two classification rely on fewer than 10
indices.

Four of the entries in Table 6 have been marked with asterisks, indicating
that the spectra are unusual. Two of these -- 2M~0739-4926 and 2M~0811-4319 --
have substantially elevated blue continua, while the blue continuum in 2M~0838-3211
is substantially depressed compared to the red part of the spectrum. A blue excess
could easily be explained by the presence of a faint background star (which is
overwhelmingly more likely to be bluer than the UCD), but a reason for the 
absence of blue light in 2M~0838-3211 is less obvious. The spectra of 2M~0819-4706
shows contamination by diffuse galactic H~II emission. 

A graphical comparison between TF and MIF spectral classifications is made in
Fig. 9, for all the stars in Tables 3-6. The are no global
systematic differences between the two classification schemes. There are three
objects with classification differences larger than 1.5 subclasses -- two
of these are known to have unusual spectra, namely   
2M~0141-4633 (TF type L2) and 2M~1126-5003 (TF type L6). The former is known
to be a young, low gravity UCD (Kirkpatrick et al. 2006, 2008; Cruz et al. 2009)
while the latter has an unusual spectrum, ascribed to the presence of condensates
in its photosphere (Burgasser et al. 2008). The third discrepant point in Fig. 9
is due to 2M~0819-4706, mentioned in the preceding paragraph.

\section*{\large \bf 5 H$\alpha$ EMISSION}

Many papers have been published on ``activity" in UCDs, the term
usually meaning ``the presence of H$\alpha$ emission". A sampling 
is Schmidt et al. (2007), Reiners \& Basri (2008), 
Berger et al. (2009), Lee et al. (2010),
Berger et al (2010), Stelzer et al. (2012), Schmidt et al. (2015), and 
Pineda et al. (2016). References to other (particularly earlier) work can
also be found in these references.
Although it is usually assumed that activity is magnetic, 
accretion is another possibility (e.g. Koen 2008, Scholz et al. 2009).

Out of the 81 UCDs with SALT spectra 28 showed measurable H$\alpha$ emission.
Equivalent widths for the individual spectra are listed in the last columns
of Tables 3-6, and a histogram of the results is plotted in Fig. 10. 
Three of the UCDs observed on multiple occasions showed intermittent emission,
while emission was either consistently absent or present in the remaining 25
objects observed more than once.
 
The binary UCD 2M~1520-4422, responsible for the outlying point in Fig. 10, 
was observed at two epochs. The two pairs
of spectra acquired are plotted in Fig. 11. Remarkable H$\alpha$ emission
(EWs of 135 and 157 \AA) can be seen in spectra obtained in June 2014.
An interesting point is that there is no apparent enhancement of the
blue continuum in these earlier spectra, as compared to the later pair 
(see Fig. 12). This
contrasts with e.g. the flaring behaviour observed by Liebert et al. (1999)
in the M9.5 object 2MASS~J0149090+295613. It is also noteworthy that
there is no trace at all of emission in the later spectra:
Phan-Bao et al. (2008) also quote an EW for 2M~1520-4433 smaller than 4 \AA\ .
Notably, no other emission features aside from  
H$\alpha$ are present in any of the spectra. 

The different responses to flaring seen in the continua
of 2M~1520-4422 and 2M~0149+2956 is interesting. Note that 2M~1520-4422 is not unique
in showing no continuum enhancement -- compare fig. 11 in Schmidt et al. (2007),
which shows three spectra of 2MASS~J10224821+5825453 (type L1). Although H$\alpha$ 
emission EWs varied from 24 to 128 \AA\ the continua appear unaffected. It is tempting
to think that flares in UCD of spectral types L are not accompanied
by continuum increases, while the converse holds for earlier spectral types.
However, ``white light" flares have also been seen in the L1 dwarf
WISEP J190648.47+401106.8 (Gizis et al. 2013) and L0 dwarf 
SDSS~J053341.43+001434.1 (Schmidt et al. 2016). Conditions which determine the 
presence or absence of increases in continuum emission in L dwarfs clearly deserve
more study.

Another point which deserved discussion 
is that the $J$ magnitudes of the two component of 2M~1520-4422 differ by about
1.15 mag (Burgasser et al. 2007). Given the difference of 2-3 in spectral subclasses
(Kendall et al. 2007, Burgasser et al. 2007), the blue flux of the fainter companion
could be about an order of magnitude below that of the brighter object. This implies
that a blue continuum enhancement due to a flare in the fainter companion may be 
difficult to detect.   

The ratio of H$\alpha$ luminosity $L_{H\alpha}$ to bolometric luminosity
$L_{\rm bol}$ is a standard measure of the level of activity. A 
straightforward way of calculating it is to use the equation
$$L_{H\alpha}/L_{\rm bol}=\chi {\rm EW(H\alpha)} \; .$$
The factor $\chi$, which is a function of spectral class, was introduced by
Walkowicz et al. (2004), and is conveniently tabulated for UCDs in
Schmidt et al. (2014). The latter paper gives a mean value of $\chi=2.18 \times
10^{-6}$ for both L1 and L2 subclasses, leading to
$\log~(L_{H\alpha}/L_{\rm bol})=-3.53, -3.47$ for the equivalent widths
quoted above.
The above assumes that the brighter A component of 2M~1520-4422 is responsible for
the emission; if it is the later B component (L4-L4.5), $\chi \approx 1.2 \times 10^{-6}$ 
and $\log~(L_{H\alpha}/L_{\rm bol})=-3.79, -3.72$.

An alternative activity measure is the ratio of H$\alpha$ and bolometric
fluxes (e.g. Schmidt et al. 2007). Burgasser et al. (2007) estimate NIR
magnitudes of the A and B components of 2M~1520-4422 as $J=13.55$, $K_S=12.27$;
$J=14.70$, $K_S=13.22$ respectively. Using bolometric corrections for
spectral types of L1.5 and L4.5 interpolated in the
tables of Schmidt et al. (2014), mean apparent bolometric magnitudes of 15.49 and
16.50 are obtained for the two components. From the relation
$$m_{\rm bol}=-2.5\log F_{\rm bol}-11.48$$
and H$\alpha$ line fluxes of 1.08 and 1.20$\times 10^{-14}$ erg cm$^{-2}$ s$^{-1}$,
$\log (F_{H\alpha}/F_{\rm bol})=-3.18, -3.13$ if the emission is associated with
the brighter component, or $\log (F_{H\alpha}/F_{\rm bol})=-2.77, -2.73$ if
it originates in the fainter component. 
 
These activity strengths are far in excess of any seen at these spectral types
in the substantial sample (181 L dwarfs) discussed by
Schmidt et al. (2015) (see their figs. 6 and 7). The only L dwarfs for which
comparable activity levels have, to the authors' knowledge, been reported are
2MASS~J1315309-264951 (Gizis 2002, Hall 2002, Burgasser et al. 2011, and
references therein) and 2MASS~J10224821+5825453 (Schmidt et al. 2007).
Spectral classification of 2M~1315-2649 is L5e (indicating sustained,
rather than flaring emission). The discovery papers reported
H$\alpha$ equivalent widths $\sim 100$ \AA\ and activity measures 
$\log~(L_{H\alpha}/L_{\rm bol})$
in the range -4.1 to -3.9. In the case of 2M~1022+5825 (L1), an H$\alpha$ EW 
of 128 \AA\ was observed on one occasion, with lower levels of emission on
two other nights (Schmidt et al. 2007).

A lesson imparted by the above is that it may be risky to draw
conclusions about activity levels from the information currently available:
given the remarkable difference between the two sets of spectra in Fig. 10,
several visits to any one target may be required to elicit the range
of its H$\alpha$ emission.

\section*{\large \bf 6 CONCLUSIONS}
The primary results of this paper are
\begin{itemize}
\item[(i)]
Quantification of the uncertainty in spectral classification,
by means of robust variance-comparison statistics, was proposed. 
\item[(ii)]
The full collection of SALT spectra was used to evaluate classification
by twenty different spectral indices. Half of these generally have
large standard errors; this includes the alkali-metal gravity-sensitive
indices.
\item[(iii)]
An alternative classification method by spectral indices was introduced,
namely the median of all spectral types implied by all accurately 
determined indices. 
\item[(iv)]
There is generally good agreement between the results of template fitting
of SALT spectra, and published classifications. In a number of instances
revised spectral types have been proposed -- typically when fits are
unambiguous (one-member confidence sets), with closely similar MIF 
classifications.    
\item[(v)]
First spectral classifications of 32 UCD candidates were presented in Table 
6. The great majority of classifications are either unambiguous, or
of types M8-M9, which often suggests a class of M8.5.
\item[(vi)]
Measurable H$\alpha$ emission was seen in 28 of the UCDs. For most of
the objects observed at multiple epochs the emission is either 
consistently present, or consistently absent.  
\item[(vii)]
One set of spectra of 2MASS~1520-4422 showed H$\alpha$ flares with equivalent 
widths in excess of 130 \AA . Interestingly, spectra at a different epoch 
had no sign of even low level emission.
\end{itemize}

It should be emphasized that although classification based on 
continuum matching is favoured in this paper, considerable additional
information is available from detailed examination of spectral features, 
as described, for example, by spectral indices. Examples are sensitivity 
to gravity (Kirkpatrick et al. 2008, Cruz et al. 2009, 
Mart\'{\i}n et al. 2010), metallicity (L\'epine, Rich \& Shara 2007;
Burgasser 2008, West et al. 2011) and the presence of 
atmospheric condensates (Burgasser
et al. 2008).

\section*{\large \bf ACKNOWLEDGMENTS}
CK, BM and PV acknowledge funding from the South African
National Research Foundation. 
Extensive use of the Simbad data base and Vizier 
catalogue service is gratefully acknowledged.
The authors thank the anonymous referee for helpful comments and 
suggestions that have allowed them to clarify some aspects of the paper.
The observations reported in this paper were obtained with the
Southern African Large Telescope (SALT) under programs
2014-1-RSA-003, 2014-2-MLT-003 and 2015-1-MLT-003.

{\bf Table 1.}\  \ Log of SALT-RSS observations.
\\\\

   \begin{tabular}{llr}
   Name & Date & Exposure \\
   (2MASS~J) & (mm/dd/yyyy) & (s)\\
   \hline
   10484281$+$0111580 & 04/16/2014 & 2$\times$600\\
   14540797$-$6604476 & 04/16/2014 & 2$\times$600\\
   15074769$-$1627386 & 04/19/2014 & 2$\times$600\\
   19165762$+$0509021 & 05/03/2014 & 800         \\    
   16553529$-$0823401 & 05/03/2014 & 600         \\      
   17502484$-$0016151 & 05/11/2014 & 2$\times$600\\
   07123786$-$6155528 & 05/12/2014 & 2$\times$600\\
   08472872$-$1532372 & 05/12/2014 & 2$\times$600\\
   10101480$-$0406499 & 05/12/2014 & 2$\times$600\\ 
   17453466$-$1640538 & 05/12/2014 & 2$\times$600\\ 
   22134491$-$2136079 & 05/12/2014 & 2$\times$600\\ 
   12212770$+$0257198 & 06/12/2014 & 2$\times$600\\        
   12281523$-$1547342 & 06/12/2014 & 2$\times$600\\        
   00325584$-$4405058 & 06/12/2014 & 2$\times$600\\        
   17502484$-$0016151 & 06/29/2014 & 600, 700\\    
   17453466$-$1640538 & 06/29/2014 & 2$\times$600\\        
   16553529$-$0823401   & 06/30/2014 & 600\\      
   19165762$+$0509021 & 06/30/2014 & 800\\      
   15074769$-$1627386 & 07/07/2014 &  2$\times$600\\      
   14540797$-$6604476 & 07/09/2014 &  2$\times$600\\      
   22134491$-$2136079 & 07/09/2014 &  2$\times$600\\      
   23225299$-$6151275 & 07/09/2014 &  2$\times$600\\      
   22244381$-$0158521 & 07/09/2014 &  2$\times$600\\      
   00332386$-$1521309 & 07/12/2014 & 2$\times$600\\ 
   01282664$-$5545343 & 07/12/2014 & 2$\times$600\\ 
   17054834$-$0516462 & 07/13/2014 & 2$\times$600\\ 
   13054019$-$2541059 & 07/13/2014 & 2$\times$600\\ 
   23225299$-$6151275 & 07/14/2014 & 2$\times$600\\ 
   01415823$-$4633574 & 07/14/2014 & 2$\times$600\\ 
   02235464$-$5815067 & 07/14/2014 & 600\\ 
   00242463$-$0158201  & 07/31/2014 & 600\\
   08593060$-$6605084 & 12/12/2014 & 3$\times$600 \\
   00145575$-$4844171 & 12/15/2014 & 3$\times$600 \\
   07522427$-$3925410 & 01/01/2015 & 2$\times$600\\ 
   06512977$-$1446150 & 01/02/2015 & 2$\times$600\\ 
   12342370$-$5104354 & 01/02/2015 & 2$\times$600\\ 
   08115730$-$4319238 & 01/04/2015 & 2$\times$600\\
   09420802$-$3758418 & 01/04/2015 & 2$\times$600\\
   11032796$-$5933001 & 01/04/2015 & 2$\times$600\\
\hline
\end{tabular}
\pagebreak

{\bf Table 1.}\  \ continued.
\\\\

   \begin{tabular}{llr}
   Name & Date & Exposure \\
   (2MASS~J) & (mm/dd/yyyy) & (s)\\
   \hline  
   09474621$-$3810043 & 01/05/2015 & 2$\times$600\\
   10192447$-$2707171 & 01/05/2015 & 2$\times$600\\
   10503781$-$4517010 & 01/05/2015 & 2$\times$600\\
   12405746$-$6549554 & 01/05/2015 & 2$\times$600\\
   08383219$-$3211406 & 01/06/2015 & 2$\times$600\\
   08274661$-$1619256 & 01/06/2015 & 2$\times$470\\
   11263991$-$5003550 & 01/06/2015 & 2$\times$600\\
   11312945$-$6446032 & 01/06/2015 & 2$\times$600\\
   11224462$-$6533161 & 01/06/2015 & 2$\times$600\\
   07595440$-$2117123 & 01/07/2015 & 2$\times$600\\
   06482289$-$2916280 & 01/15/2015 & 2$\times$600\\
   08193434$-$4706133 & 01/30/2015 & 2$\times$600\\
   08143545$-$4020492 & 02/01/2015 & 2$\times$600\\  
   11301046$-$5759419 & 02/04/2015 & 2$\times$600\\
   07410404$-$0359495 & 02/07/2015 & 600\\  
   13080663$-$4925505 & 03/02/2015 & 2$\times$600\\   
   07561708$-$0715512 & 03/07/2015 & 2$\times$600\\     
   08173001$-$6155158 & 03/27/2015 & 2$\times$600\\     
   06465202$-$3244011 & 03/28/2015 & 2$\times$600\\
   07293904$-$2608578 & 03/28/2015 & 2$\times$600\\
   17502484$-$0016151 & 03/28/2015 & 2$\times$600\\
   06164933$-$1411434 & 03/29/2015 & 2$\times$600\\
   12065011$-$3937261 & 03/29/2015 & 2$\times$600\\
   07313276$-$2841575 & 03/29/2015 & 2$\times$600\\
   10565008$-$6122042 & 03/30/2015 & 2$\times$600\\
   17373472$-$3446108 & 03/31/2015 & 2$\times$600\\
   17054834$-$0516462 & 03/31/2015 & 2$\times$600\\
   13201384$-$4842117 & 03/31/2015 & 2$\times$600\\
   07390794$-$4926533 & 04/24/2015 & 2$\times$600\\
   18422444$-$1358138 & 04/30/2015 & 550, 600\\
   14263345$-$5229166 & 04/30/2015 & 600\\
   16445570$-$2618333 & 04/30/2015 & 2$\times$600\\
   16445570$-$2618333 & 05/02/2015 & 2$\times$600\\
   06482289$-$2916280 & 05/03/2015 & 2$\times$600\\
   07293904$-$2608578 & 05/03/2015 & 2$\times$600\\
   11224462$-$6533161 & 05/03/2015 & 2$\times$600\\
\hline
\end{tabular}
\pagebreak

{\bf Table 1.}\  \ continued.
\\\\

   \begin{tabular}{llr}
   Name & Date & Exposure \\
   (2MASS~J) & (mm/dd/yyyy) & (s)\\
   \hline 
 
   14334194$-$5148037 & 05/04/2015 & 2$\times$600\\
   08193434$-$4706133 & 05/24/2015 & 3$\times$600\\
   10192447$-$2707171 & 05/25/2015 & 2$\times$600\\
   17502484$-$0016151 & 06/18/2015 & 2$\times$600\\
   15200224$-$4422419 & 06/22/2015 & 2$\times$600\\
   00065794$-$6436542 & 07/01/2015 & 600\\
   21501324$-$6610366 & 07/01/2015 & 2$\times$600\\
   00311925$-$3840356 & 07/02/2015 & 2$\times$600\\
   17502484$-$0016151 & 07/03/2015 & 2$\times$600\\
   15200224$-$4422419 & 07/03/2015 & 2$\times$600\\
   15485834$-$1636018 & 07/06/2015 & 2$\times$600\\
   00065794$-$6436542 & 07/06/2015 & 2$\times$1200\\
   17072529$-$0138093 & 07/13/2015 & 2$\times$600\\
   20131084$-$1242452 & 07/13/2015 & 600\\
   21501324$-$6610366 & 07/15/2015 & 2$\times$600\\
   20131084$-$1242452 & 07/19/2015 & 600\\
   13080663$-$4925505 & 07/22/2015 & 229, 600\\
   12191303$-$5021426 & 07/28/2015 & 2$\times$600\\
   14324269$-$5534247 & 07/28/2015 & 2$\times$600\\
   22551861$-$5713056 & 07/28/2015 & 2$\times$600\\
   00311925$-$3840356 & 07/28/2015 & 2$\times$560\\
   17072529$-$0138093 & 08/01/2015 & 2$\times$600\\
   15485834$-$1636018 & 08/01/2015 & 2$\times$600\\
   01165283$-$6455570 & 08/01/2015 & 2$\times$600\\
   17054834$-$0516462 & 08/02/2015 & 2$\times$600\\
   01165283$-$6455570 & 08/06/2015 & 2$\times$600\\
   22521073$-$1730134 & 08/06/2015 & 2$\times$600\\
   23081134$-$2722001 & 08/06/2015 & 2$\times$600\\
   15410782$-$6026051 & 08/12/2015 & 3$\times$600\\
   18052143$-$0733179 & 08/12/2015 & 2$\times$600\\
   14324269$-$5534247 & 08/14/2015 & 2$\times$600\\
   16445570$-$2618333 & 08/14/2015 & 2$\times$600\\
   01333248$-$6314415 & 08/14/2015 & 2$\times$600\\
   14334194$-$5148037 & 08/15/2015 & 2$\times$600\\  
   19265883$-$0844206 & 08/15/2015 & 2$\times$530\\
   14263345$-$5229166 & 08/16/2015 & 2$\times$600\\
   15230657$-$2347526 & 08/16/2015 & 2$\times$600\\

   \hline
\end{tabular}

\pagebreak

{\bf Table 1.}\  \ continued.
\\\\

   \begin{tabular}{llr}
   Name & Date & Exposure \\
   (2MASS~J) & (mm/dd/yyyy) & (s)\\
   \hline  

   16533670$-$3855165 & 08/16/2015 & 2$\times$600\\
   16532340$-$6424077 & 08/17/2015 & 2$\times$600\\
   18473965$-$1856577 & 08/17/2015 & 3$\times$600\\
   18002648$+$0134566 & 08/21/2015 & 2$\times$600\\
   17275293$-$6227029 & 08/24/2015 & 2$\times$600\\
   23211254$-$1326282 & 08/24/2015 & 500, 600\\
   15410782$-$6026051 & 09/08/2015 & 350, 2$\times$600\\
   00145575$-$4844171 & 10/01/2015 & 2$\times$600\\
   01333248$-$6314415 & 10/01/2015 & 2$\times$600\\
   07293904$-$2608578 & 10/01/2015 & 2$\times$600\\
   18450541$-$6357475 & 10/02/2015 & 2$\times$600\\
   06164933$-$1411434 & 10/03/2015 & 2$\times$600\\
   02304498$-$0953050 & 10/03/2015 & 3$\times$600\\
   07230144$-$1616209 & 10/05/2015 & 2$\times$600\\
   06400355$-$1449104 & 10/06/2015 & 2$\times$600\\
   06512977$-$1446150 & 10/06/2015 & 2$\times$600\\
   \hline
\end{tabular}

%

\vspace*{2cm}

{\bf Table 2.}\  \ Spectral indices considered for use in classifications:
the wavelength intervals (in \AA\ ) used for calculation of the numerator and
denominator of the flux ratios are listed.
In some cases the inverses of indices defined in the literature
were used, in order that the ranges of numerical values 
be similar across all 22 indices. 
\\\\

\begin{tabular}{llcccllcc}

No. & Name & Numerator & Denominator && No. & Name & Numerator 
& Denominator\\ 
& & & && & & \\
1& TiO1   & 6718-6723 & 6703-6708 && 10&  K-b    & 7550-7570 & 7690-7710 \\
2& CaH2   & 6814-6846 & 7042-7046 && 11&  VO7912 & 7900-7980 & 8400-8420 \\
3& CaH3   & 6960-6990 & 7042-7046 && 12&  Na8190 & 8140-8165 & 8173-8210 \\
4& TiO2   & 7058-7061 & 7043 7046 && 13&  Na-a   & 8153-8163 & 8178-8188 \\
5& TiO-a  & 7058-7073 & 7033-7048 && 14&  Na-b   & 8153-8163 & 8190-8200 \\
6& TiO3   & 7092-7097 & 7042-7046 && 15&  TiO8440& 8440-8470 & 8400-8420 \\
7& TiO5   & 7126-7135 & 7042-7046 && 16&  CrH-a  & 8580-8600 & 8621-8641 \\
8& TiO4   & 7130-7135 & 7042-7046 && 17&  FeH-a  & 8660-8680 & 8700-8720 \\
9& VO7434 & 7430-7470 & 7550-7570 &&   &       &           &            \\
& & & && & & \\
18 &CaH1   & 6380-6390 & 6345-6355, 6410-6420 && & && \\
19 &Rb-a   & 7795-7805 & 7775-7785, 7815-7825 && & & &\\
20& Rb-b   & 7943-7953 & 7923-7933, 7963-7973 && & && \\
21&VO-b   & 7960-8000 & 7860-7880, 8080-8100 && & && \\
22&Cs-a   & 8516-8526 & 8496-8506, 8536-8547 && & & &
\end{tabular}

\pagebreak

{\bf Table 3.}\  \ Instances where SALT TF classifications differ by two or more
subclasses from types given in the Simbad database. In some cases revised spectral
classifications are proposed (penultimate column). 
Non-zero H$\alpha$ emission equivalent widths are listed in the last column.
\\\\

\begin{tabular}{lllclllc}

 Name & Simbad & TF & Confidence Set && MIF & Final & EW(H$\alpha$)\\
 & & & & & & & (\AA)\\
 & & & & & & & \\
    2MASS~J01282664-5545343   &   L1 &  L3 &  L3  && L2.9 & L3 &\\
    2MASS~J02235464-5815067   &   L0 &  L2 &  L1 -  L2  && L2.1 & L1.5 & \\
    2MASS~J02304498-0953050   &   L6 &  L2 &  L1 -  L3  && L2.9 & L2 & \\ 
    2MASS~J15230657-2347526   &   L2.5 &  L0 &  L0   &&  M9.8 & L0  &  \\
    2MASS~J15485834-1636018   &   L2 &  L0 &  L0 && M9.9 & L0  &  \\ 
                              &   L2 &  L0 &  L0   &&  L0.0 &   & \\ 
    2MASS~J16553529-0823401   &    M7 &   M9 &   M7 - M9  &&  M7.9 & &9  \\
                              &    M7 &   M9 &   M8 - M9  &&  M7.9 & &6  \\
    2MASS~J22521073-1730134   &   T0 & L5 & L5   &&L5.8 &L5   & \\
    2MASS~J22551861-5713056   &   L5.5 &  L3 &  L3  && L3.9 &  & \\
\end{tabular}

\vspace*{2cm}

{\bf Table 4.}\  \ Instances where SALT TF classifications differ by 1.5
subclasses from types given in the Simbad database. In some cases revised spectral
classifications are proposed (penultimate column). 
Non-zero H$\alpha$ emission equivalent widths are listed in the last column.
\\\\

\begin{tabular}{lllclllc}

 Name & Simbad & TF & Confidence Set && MIF & Final & EW(H$\alpha$)\\
 & & & & & & & (\AA) \\
 & & & & & & & \\
    2MASS~J06512977-1446150   &     M7.5 &   M9 &   M8 -   M9 &&   M8.3 & M8.5&25 \\   
                              &     M7.5 &   M9 &   M8 -   M9 &&   M8.0 & &10\\
    2MASS~J08143545-4020492   &     M7-M8 &   M9 &   M8 -   M9 &&   M8.7& M8.5 &\\   
    2MASS~J08472872-1532372   &    L1.5 &  L3 &  L2 -  L3 &&  L3.2 & &\\
    2MASS~J13080663-4925505   &     M8.5 &   M7 &   M7  &&   M7.4 & M7 &5 \\   
                              &     M8.5 &   M7 &   M7  &&   M7.3 &  &4 \\
    2MASS~J14540797-6604476   &    L3.5 &  L5 &  L5 &&  L5.5 & L5 & \\ 
    2MASS~J15200224-4422419   &    L4.5 &  L3 &  L1 - L3 && L2.3 &  & 146\\
                              &    L4.5 &  L3 &  L1 - L3 && L2.2 & &\\
    2MASS~J18450541-6357475   &     M8.5 &  L0 &  L0  &&   M9.1 & M9.5 & 5\\
    2MASS~J22244381-0158521   &    L4.5 &  L6 &  L4 -  L6 &&  L6.2 &  &  \\
    2MASS~J23081134-2722001   &    L1.5 &  L3 &  L1 -  L3 &&  L3.2 & &10

\end{tabular}

\pagebreak

{\bf Table 5.}\  \ UCDs for which SALT TF classifications differ by less than 1.5
subclasses from types given in the Simbad database. In some cases revised spectral
classifications are proposed (penultimate column). 
Non-zero H$\alpha$ emission equivalent widths are listed in the last column.
\\\\

\begin{tabular}{lllclllc}

 Name & Simbad & TF & Confidence Set && MIF & Final & EW(H$\alpha$)\\
 & & & & & & & (\AA)\\
 & & & & & & & \\
     2MASS~J00065794-6436542    &    M9    &     L0     &   M8 - L0      &&     M9.4  &M9.5 &\\
     2MASS~J00145575-4844171    &   L2.5   &      L3    &    L3     &&      L4.0 &L3& \\
                                &   L2.5   &      L3    &    L1 - L3     &&      L2.5 & &\\    
     2MASS~J00242463-0158201     &   M9.5  &       L0   &     L0     &&       L0.4  & L0&\\
     2MASS~J00311925-3840356    &   L2     &    L1      &  L1 - L3       &&    L1.5 & &  \\ 
                                &   L2     &    L3      &  L3       &&    L2.1  &  &\\  
     2MASS~J00325584-4405058     &   L0     &    L0      &  L0 - L2       &&    L0.3 &  &\\ 
     2MASS~J01165283-6455570    &   L1     &    L1      &  L0 - L1       &&    L0.8 & &  \\ 
     2MASS~J01333248-6314415    &    M8    &     M7     &  M7        &&  M7.3 &M7 &3.5      \\    
                                &    M8    &     M7     &  M7        &&  M7.2 & & 5      \\ 
     2MASS~J01415823-4633574     &   L2     &    L2      &  L1 - L2       &&    L0.1  & &  \\ 
     2MASS~J07123786-6155528     &   L1     &    L2      &  L2 - L3       &&    L3.0  &L2 &15\\ 
     2MASS~J08274661-1619256     &    M8    &     M9     &   M8 - M9      &&      M8.8 &M8.5 & \\   
     2MASS~J10192447-2707171     &    M9.5  &       L0   &     L0    &&       M9.5 &L0 & 5\\ 
                                 &    M9.5  &       L0   &     L0     &&       L0.0 & &9 \\ 
     2MASS~J10484281+0111580     &   L1     &    L1      &  L1        &&    L0.9  &  & \\ 
     2MASS~J10503781-4517010     &    M9    &     L0     &   L0      &&      M9.1 &L0 &10 \\
     2MASS~J11263991-5003550     &   L5     &    L6      &  L3 - L6       &&    L3.8 &&   \\ 
     2MASS~J12065011-3937261     &   L2     &    L2      &  L2 - L3       &&    L2.4 &L2.5 &  \\      
     2MASS~J12212770+0257198     &   L0.5   &      L0    &    L0     &&      L0.2 &L0 &2.5 \\ 
     2MASS~J12281523-1547342     &   L5     &    L5      &  L5 - L6       &&    L6.3  &L5.5 & \\ 
     2MASS~J13054019-2541059     &   L2     &    L3      &  L3       &&    L3.1  &L3 & \\ 
                                 &   L2     &    L3      &  L3        &&    L3.2 &L3  & \\ 
     2MASS~J14263345-5229166    &    M8    &     M7     &  M7        &&  M7.1  &  M7  & \\ 
                                &    M8    &     M7     &  M7        &&  M7.4 &  &    \\ 
     2MASS~J14334194-5148037    &    M6.5  &       M7   &    M7      &&    M7.1 &M7&20     \\ 
                                &    M6.5  &       M7   &    M7      &&    M7.3 &  &16   \\       
     2MASS~J15074769-1627386     &   L5     &    L5      &  L5       &&    L5.7 &  &  \\ 
                                 &   L5     &    L6      &  L5 - L6       &&    L5.2 & &   \\ 
     2MASS~J17054834-0516462     &   L1     &    L1      &  L1        &&    L1.2  & &  \\ 
                                 &   L1     &    L1      &  L1       &&    L1.5   & & \\ 
                                 &   L1     &    L1      &  L1        &&    L1.0 & &    \\ 
     2MASS~J17072529-0138093    &   L0.5   &      L1    &    L1     &&      L1.0 &L1 &  \\       
     2MASS~J17453466-1640538     &   L1.5   &      L1    &    L1     &&      L0.9 &L1 & \\ 
                                 &   L1.5   &      L1    &    L1    &&      L1.1  & &\\ 
\end{tabular}

\pagebreak
{\bf Table 5.}\  \ continued.
\\\\

\begin{tabular}{lllclllc}

 Name & Simbad & TF & Confidence Set && MIF & Final & EW(H$\alpha$)\\
 & & & & & & & (\AA)\\
 & & & & & & & \\
     2MASS~J17502484-0016151     &   L4.5   &      L5    &    L5     &&      L5.3 &L5 & \\ 
                                 &   L4.5   &      L5    &    L5     &&      L5.2 & & \\ 
                                 &   L4.5   &      L5    &    L5     &&      L5.7  && \\  
                                 &   L4.5   &      L5    &    L5     &&      L6.0 & &  \\  
                                 &   L4.5   &      L5    &    L5     &&      L5.7 & &  \\ 
     2MASS~J18002648+0134566    &   L7.5   &      L7    &    L7 - L8     &&   L7.3 &  & \\    
     2MASS~J19165762+0509021     &    M8    &      M9    &     M8 - M9    &&  M8.9 &M8.5&5 \\ 
                                &    M8    &      M9    &     M8 - M9    &&  M9.0 & &8\\  
     2MASS~J20131084-1242452    &   L1.5   &      L1    &    L1 - L2     &&   L0.8 & & \\ 
                                 &   L1.5   &      L2    &    L1 - L3     &&   L1.9  & & \\ 
     2MASS~J21501324-6610366    &    M9    &     L0     &   M8 - L0      &&   M9.5 &&    \\      
                                  &    M9    &     L0     &   M8 - L0      &&   M9.5 &&    \\ 
     2MASS~J22134491-2136079     &   L0     &    L1      &  L0 - L1       &&    L0.7 &L0.5 &   \\ 
     2MASS~J23211254-1326282    &   L1     &    L1      &  L1 - L2       &&    L1.0  & &   \\
     2MASS~J23225299-6151275     &   L2     &    L3      &  L2 - L4       &&    L1.5 & L2.5 &  \\
                               &   L2     &    L2      &  L1 - L3       &&    L2.9  &&
\end{tabular}

\vspace*{1.8cm}

{\bf Table 6.}\  \ New classifications. TF and MIF types are in columns 3 and 5,
with final classifications in column 7. Column 2 gives the date of observation
and column 6 the number of spectral indices used to derive the MIF type. 
Non-zero H$\alpha$ emission equivalent widths are listed in the last column.
See the text for remarks about objects marked with an asterisk.
\\\\

\begin{tabular}{lclcllccc}

 Name & Date &  TF & Confidence Set && MIF & No. indices & Final & EW(H$\alpha$)\\
 &(DD MM YY) & & & & & & & (\AA)\\
 & & & & & & & &\\
    2MASS~J06164933-1411434   & 29 03 15    &     M8 &   M8 -  L0  &  &    M8.6  &   14 &M9-L0 & \\
                              & 03 10 15    &     L0  &   L0    &  &   L0.3  &  11 & &\\
    2MASS~J06400355-1449104   & 06 10 15    &     M8 &   M8 -  M9  &  &    M8.1  &   19 &M8.5 &11 \\
    2MASS~J06465202-3244011   & 28 03 15    &     M8 &   M8 -  M9  &  &    M8.8  &   20 &M8.5 &45 \\
    2MASS~J06482289-2916280   & 15 01 15    &     M8 &   M8 -  M9  &  &    M8.0  &   15 &M8 & \\
                              & 03 05 15    &     M8 &   M8 &  &   M9.1  &   19  & &15\\
    2MASS~J07230144-1616209   & 05 10 15    &    L1 &  L1 &  &   L1.0  &   20 & L1 & \\
    2MASS~J07293904-2608578   & 28 03 15    &     M9  &   M9   &  &    M8.7  &   18 &M9 & 10\\
                              & 03 05 15    &     L0  &   M8 -  L0  &  &   M9.6  &   7 & & \\
                              & 01 10 15    &     M9  &   M9    &  &   M9.4  &   19 & & 6\\
    2MASS~J07313276-2841575   & 29 03 15    &     M9  &   M8 -  M9  &  &   M9.1  &   17&M8.5&   \\
    2MASS~J07390794-4926533*   & 24 04 15    &    L1 &  L1  &  &   L0.8  &   20 &L1 &  \\
\end{tabular}
\pagebreak

{\bf Table 6}\  \ continued.
\\\\

\begin{tabular}{lclcllccc}

 Name & Date &  TF & Confidence Set && MIF & No. indices & Final & EW(H$\alpha$)\\
  &(DD MM YY) & & & & & & & (\AA)\\
 & & & & & & & &\\
    2MASS~J07410404-0359495   & 07 02 15    &     L0  &   L0    &  &   M9.9  &   20 &L0 & \\
    2MASS~J07522427-3925410   & 01 01 15    &     M9  &   M9   &  &   M9.0  &   16 &M9 &20 \\
    2MASS~J07561708-0715512   & 07 03 15    &     M8 &   M8 -  M9  &  &    M8.9  &   15 &M8.5 & \\
    2MASS~J07595440-2117123   & 07 01 15    &     M9  &   M8 -  M9  &  &    M8.1  &   9 &M8.5 &40 \\
    2MASS~J08115730-4319238*   & 04 01 15    &     M7  &   M7    &  &    M8.1  &   19 & M7 & 10 \\
    2MASS~J08193434-4706133*   & 30 01 15    &     M7  &   M7   &  &   M7.9  &   15& M7 & \\
                              & 24 05 15    &     M7  &   M7   &  &   M9.0  &   12 & & \\
    2MASS~J08383219-3211406*   & 06 01 15    &     M7  &   M7   &  &   M7.6  &   18 &M7 & \\
    2MASS~J08593060-6605084   & 12 12 14    &     M9  &   M8 -  M9  &  &   M9.0  &   17 &M9 &25 \\
    2MASS~J09420802-3758418   & 04 01 15    &     M9  &   M8 -  M9  &  &   M8.9  &   13 &M8.5 & \\
    2MASS~J09474621-3810043   & 05 01 15    &     M8 &   M8  &  &    M8.0  &   18 &M8 & \\
    2MASS~J11224462-6533161   & 06 01 15    &     M8 &   M8 -  M9  &  &    M8.9  &   13 &M8.5 &\\
                              & 03 05 15    &     M8 &   M8 -  M9  &  &    M8.4  &   19 & & \\
    2MASS~J11312945-6446032   & 06 01 15    &     M9  &   M8 -  M9  &  &   M9.2  &   12 &M8.5 & \\
    2MASS~J12191303-5021426   & 28 07 15    &     M9  &   M9   &  &   M8.1  &   19 &M9 &33 \\
    2MASS~J12342370-5104354   & 02 01 15    &     M9  &   M8 -  M9  &  &    M8.0  &   15 &M9 &15 \\
    2MASS~J12405746-6549554   & 05 01 15    &     M7  &   M7   &  &   M7.7  &   17 &M7 & \\
    2MASS~J13201384-4842117   & 31 03 15    &     M9  &   M8 -  M9  &  &   M9.0  &   15 &M9 & 10 \\
    2MASS~J14324269-5534247   & 28 07 15    &     M9  &   M8 -  M9  &  &   M9.3  &   18 &M8.5 & \\
                              & 14 08 15    &     M8 &   M8 -  M9  &  &    M8.5  &  10 & &\\
    2MASS~J15410782-6026051   & 12 08 15    &     M6  &   M6  -  M7  &  &   M6.6  &   19 &M7 & 4\\
                              & 08 09 15    &     M7  &   M7   &  &   M7.6  &   15 & & 5\\
    2MASS~J16445570-2618333   & 30 04 15    &     M8 &   M8 -  M9  &  &   M7.6  &   18 & M8.5 & 7\\
                              & 02 05 15    &     M9  &   M8 -  M9  &  &   M7.7  &   19 & & 6 \\
                              & 14 08 15    &     M7  &   M7   &  &   M7.6  &   18 & & 7\\
    2MASS~J16533670-3855165   & 16 08 15    &     M7  &   M7  -  M9  &  &   M7.6  &   20 &M8 & 9\\
    2MASS~J17275293-6227029   & 24 08 15    &     L0  &   M8 -  L0  &  &    M8.9  &   17 &L0 & \\
    2MASS~J18422444-1358138   & 30 04 15    &     M7  &   M7    &  &   M7.3  &   19 &M7 & 8 \\
    2MASS~J18473965-1856577   & 17 08 15    &     M7  &   M7    &  &   M7.0  &   17 &M7 & \\
    2MASS~J19265883-0844206   & 15 08 15    &     M7  &   M7   &  &    M8.0  &   15 &M7 &15 \\
\end{tabular}
  
\pagebreak
\begin{figure}
\epsfysize=16.0cm
\epsffile{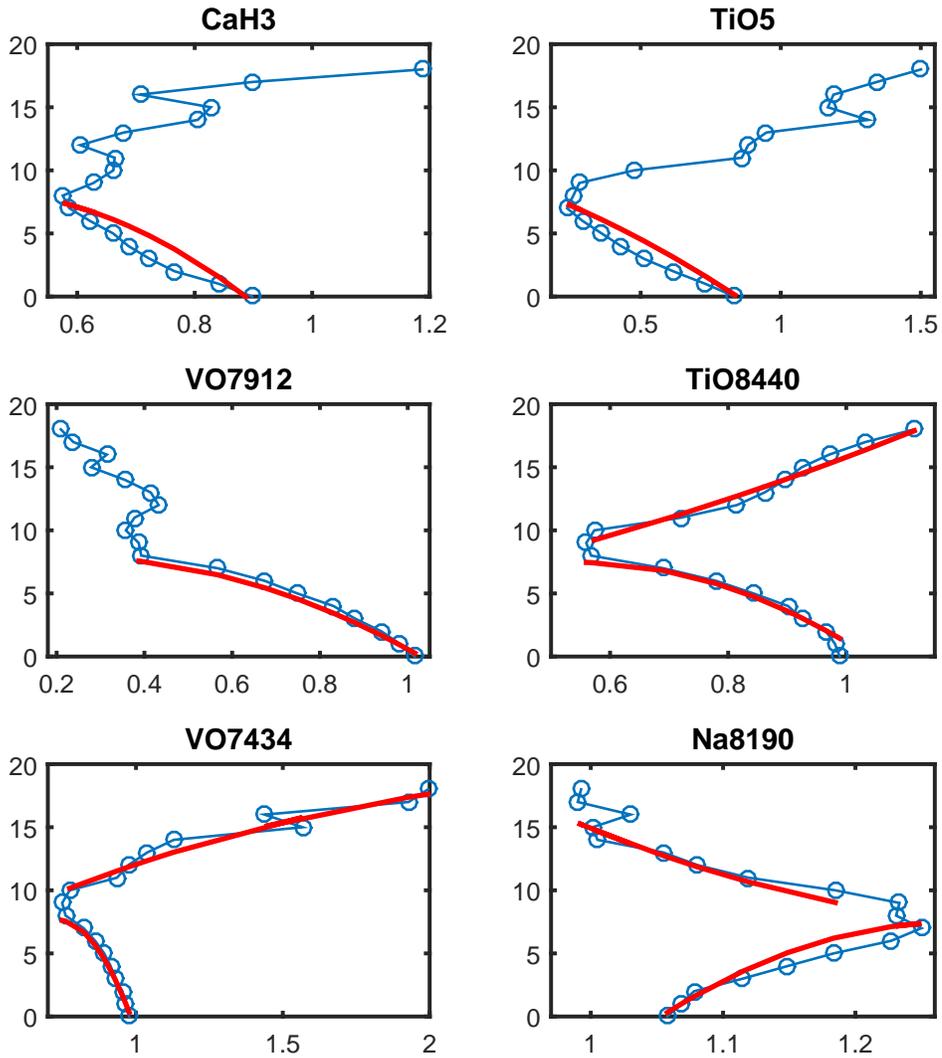}
\caption{The six spectral indices calibrated by Hawley et al. (2002) (smooth solid
lines) and the values obtained from the spectral templates (open circles). Spectral class
is plotted on the vertical axes, running from 0 (M0) to 18 (L8).}
\end{figure}

\begin{figure}
\epsfysize=16.0cm
\epsffile{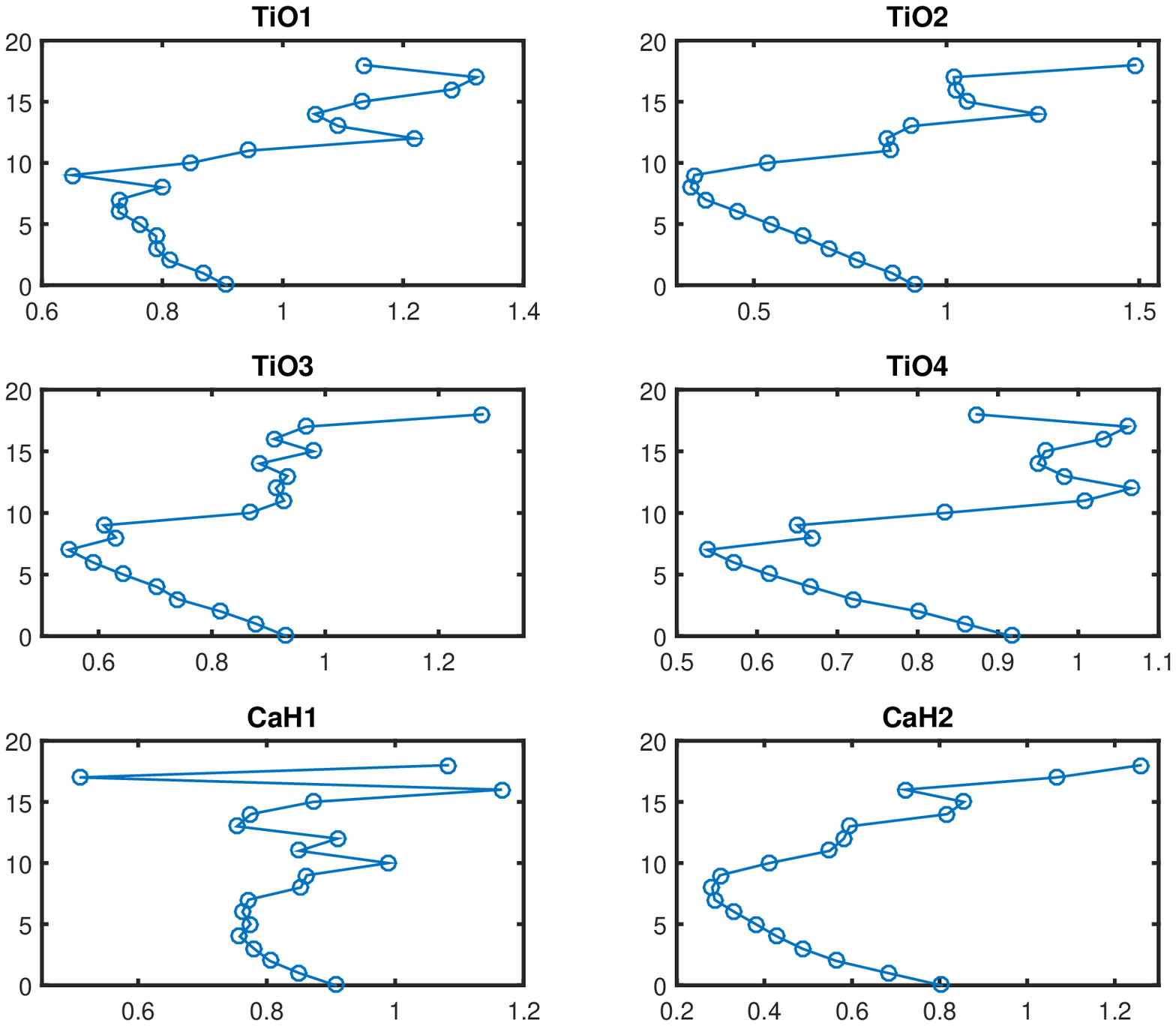}
\caption{Numerical values of six of the spectral indices defined in Table 2,
calculated from the spectral templates.}
\end{figure}

\begin{figure}
\epsfysize=16.0cm
\epsffile{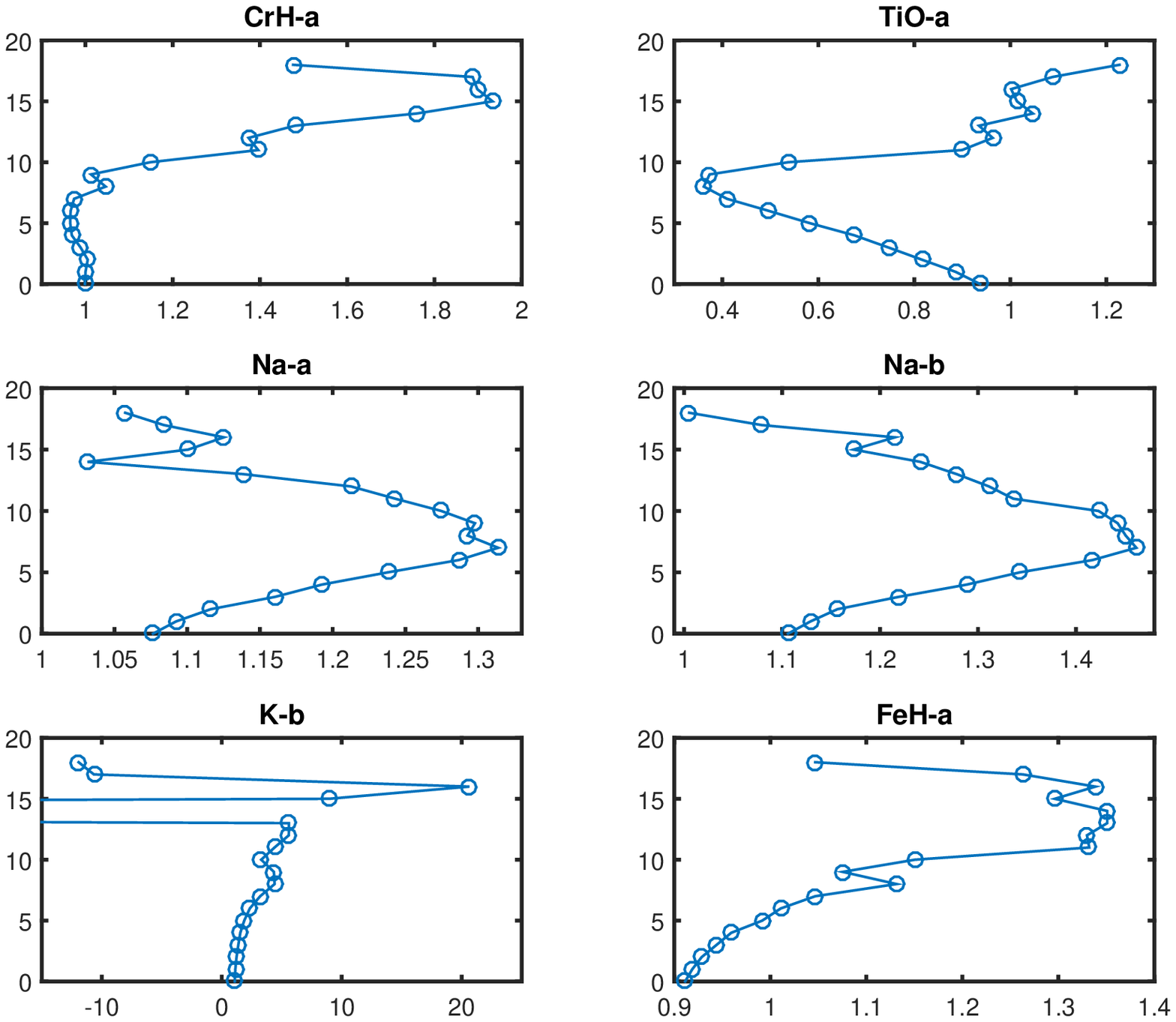}
\caption{Numerical values of six of the spectral indices defined in Table 2,
calculated from the spectral templates.}
\end{figure}

\begin{figure}
\epsfysize=11.0cm
\epsffile{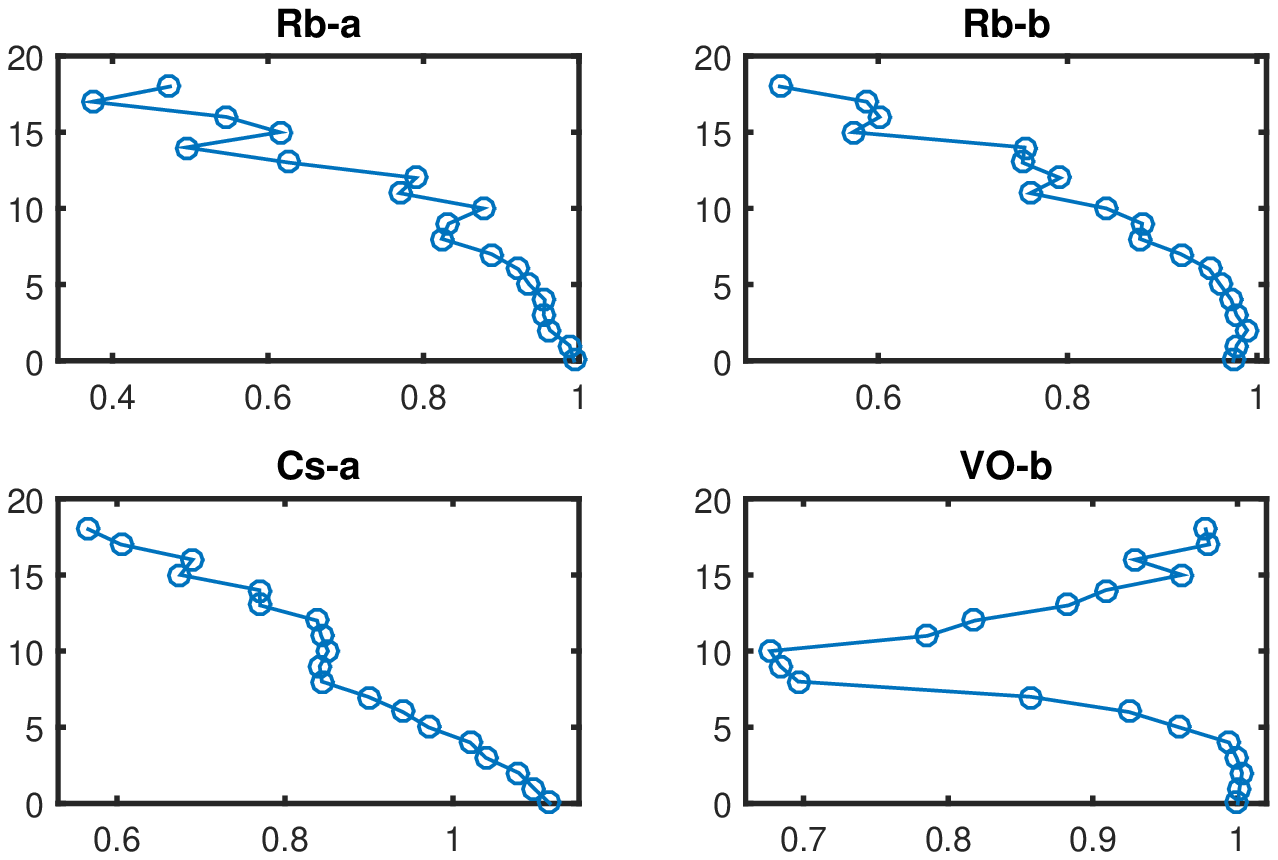}
\caption{Numerical values of four of the spectral indices defined in Table 2,
calculated from the spectral templates.}
\end{figure}

\begin{figure}
\epsfysize=7.0cm
\epsffile{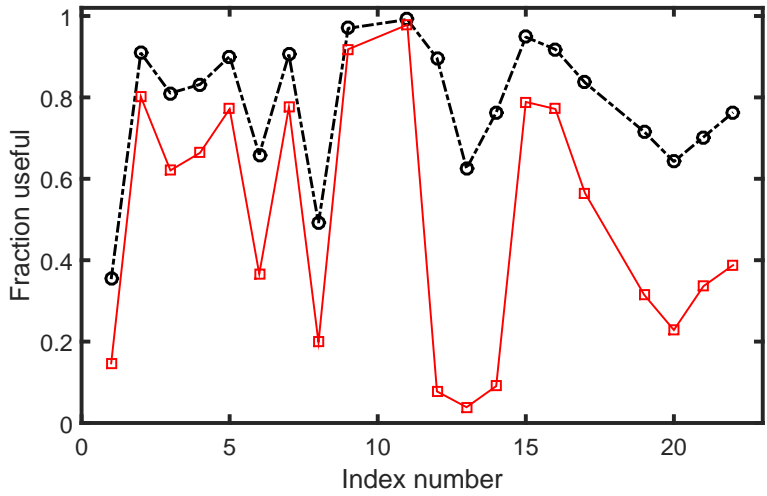}
\caption{A summary of the standard errors of 20 of the spectral indices
(see Table 1 for definitions). The K-b and CaH-1 indices have been excluded,
since these poorly calibrated for the later spectral types.
Shown are the fractions, across 232 spectra, which have estimated standard
errors smaller than 0.1 (points connected by solid lines) or smaller
than 0.2 (points connected by broken lines). Some indices are clearly
less resistant to noise.}
\end{figure}

\begin{figure}
\epsfysize=10.0cm
\epsffile{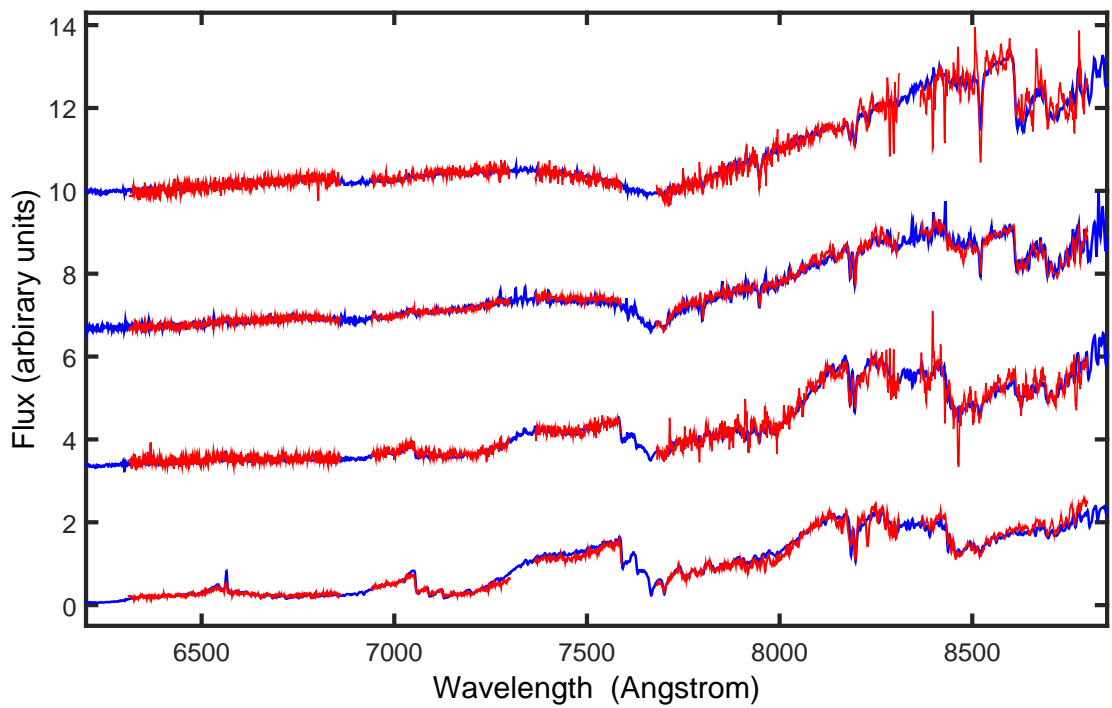}
\caption{Representative examples of the fits of SALT spectra (red) to templates (blue). 
From top to bottom,
2M~2252-1730 (L5); 2M~0128-5545 (L3); 2M~1523-2347 (L0); and
2M~1308-4925 (M7).}
\end{figure}

\begin{figure}
\epsfysize=10.0cm
\epsffile{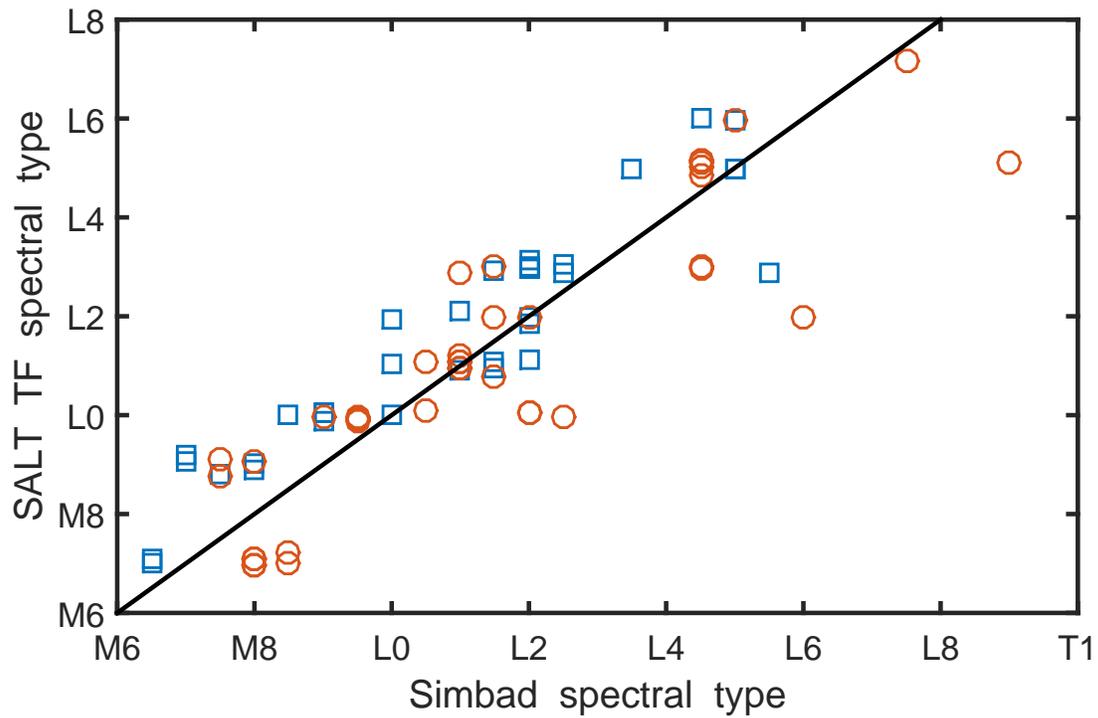}
\caption{A comparison between literature spectral types extracted from the 
Simbad database,
and those obtained from template fitting (TF) to SALT spectra. 
The line indicates equal spectral classes. Circles and squares 
respectively denote literature classification based on infrared and optical
spectra. For greater clarity the SALT spectral classes have been jittered by 
adding Gaussian random numbers with zero mean and a standard deviation of 0.1
subclasses.}
\end{figure}

\begin{figure}
\epsfysize=10.0cm
\epsffile{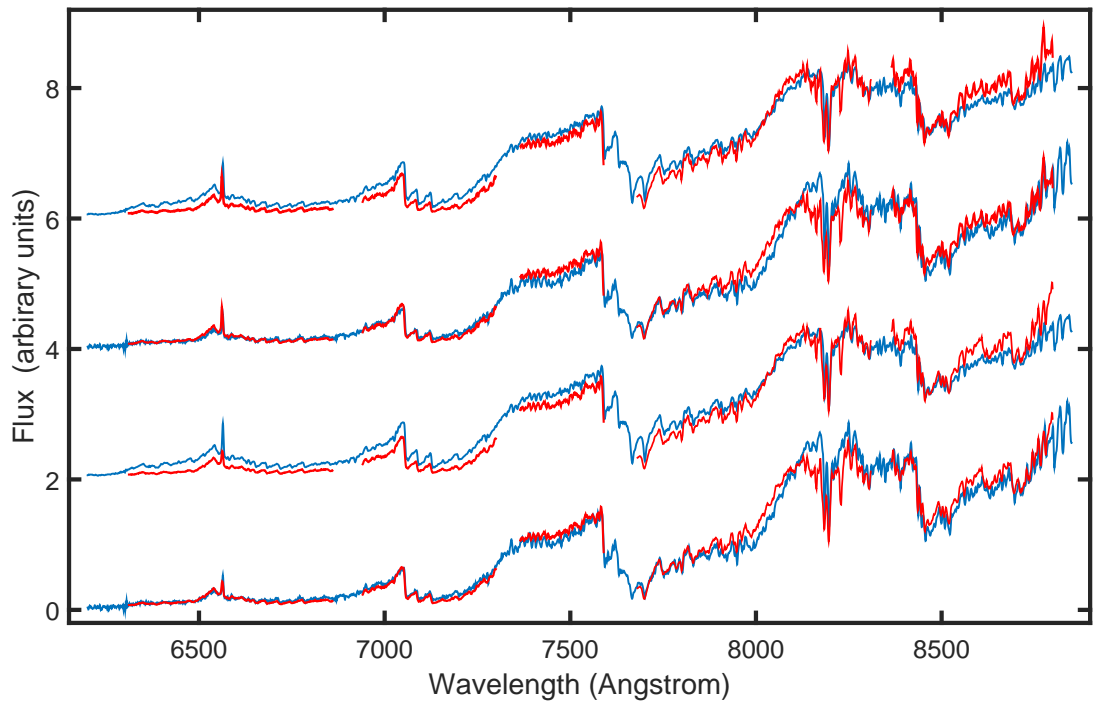}
\caption{Fits of template spectra to two SALT spectra of 2M~1655-0823 (VB~8).
The top two plots show the SALT spectrum obtained on 2014 March 12, with
M7 (top) and M9 templates. The bottom two plots show the SALT spectrum
from 2014 June 30, also with M7 (top) and M9 templates.}
\end{figure}

\begin{figure}
\epsfysize=10.0cm
\epsffile{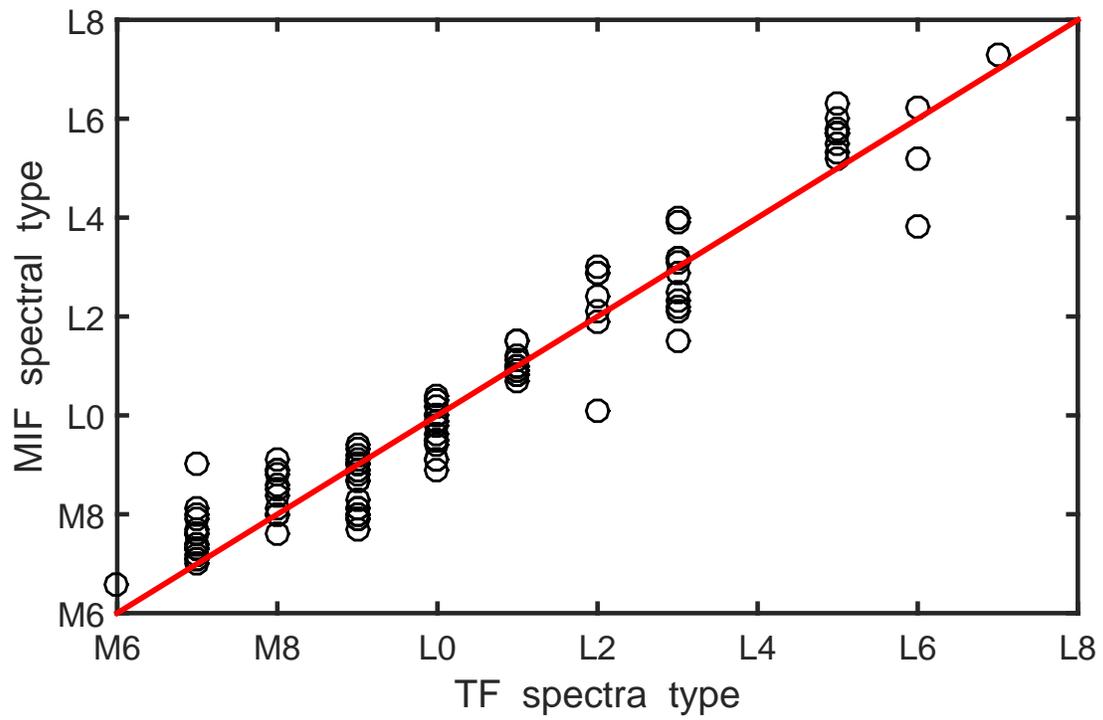}
\caption{A comparison of spectral types from template fitting (TF) and from the
the median of spectral indices (MIF), for all the stars for which SALT spectra
were obtained. The line indicates equal spectral 
classes.}
\end{figure}

\begin{figure}
\epsfysize=10.0cm
\epsffile{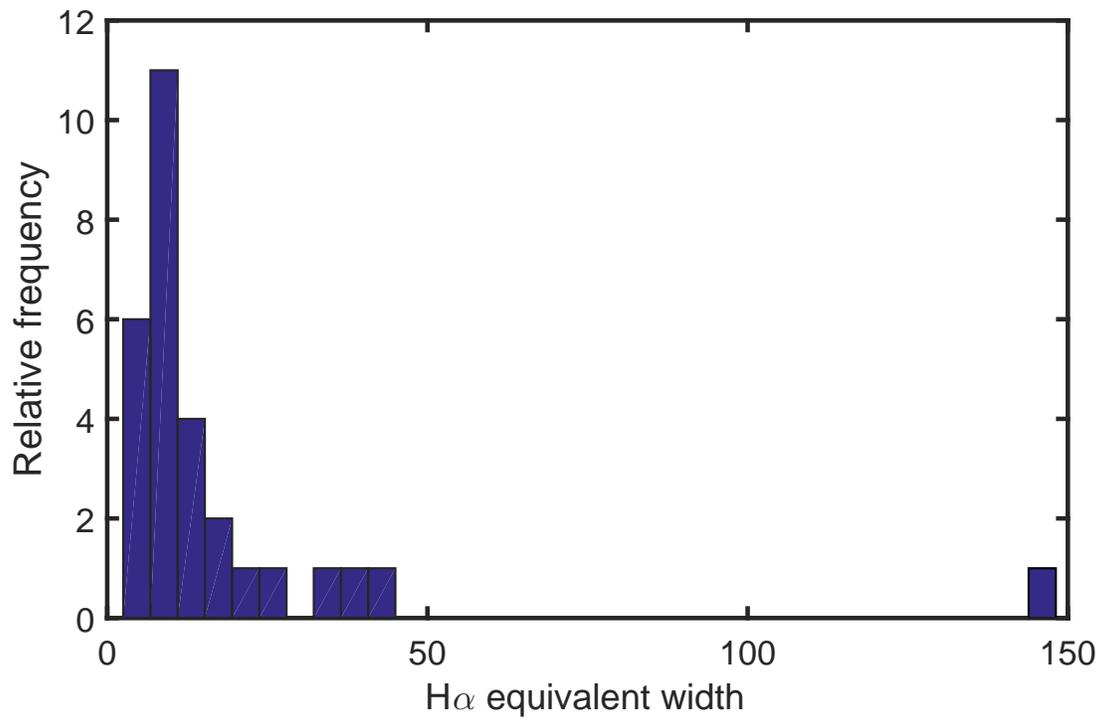}
\caption{Histogram of the non-zero H$\alpha$ EWs measured from the SALT spectra. 
Mean values are shown in those cases where objects were observed at multiple epochs.} 
\end{figure}

\begin{figure}
\epsfysize=10.0cm
\epsffile{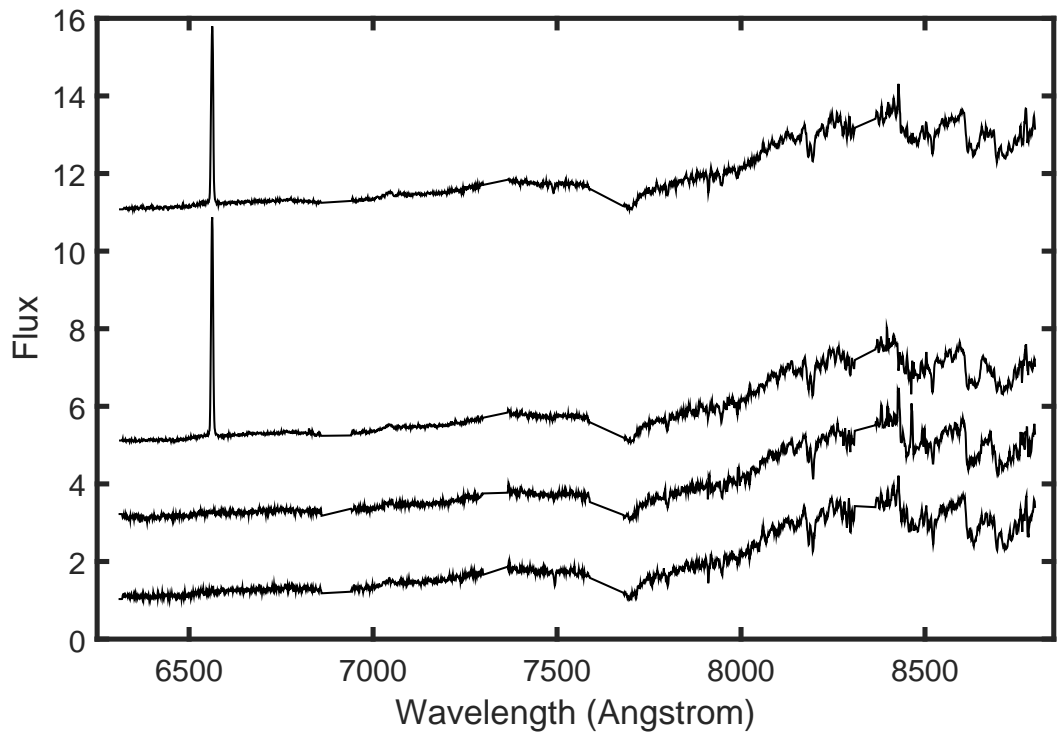}
\caption{Spectra of 2MASS~1520-4422, showing occasional very strong H$\alpha$
emission. The top two spectra were obtained in succession on 2014 June 22, the
bottom two in succession on 2014 August 3.}
\end{figure}

\begin{figure}
\epsfysize=10.0cm
\epsffile{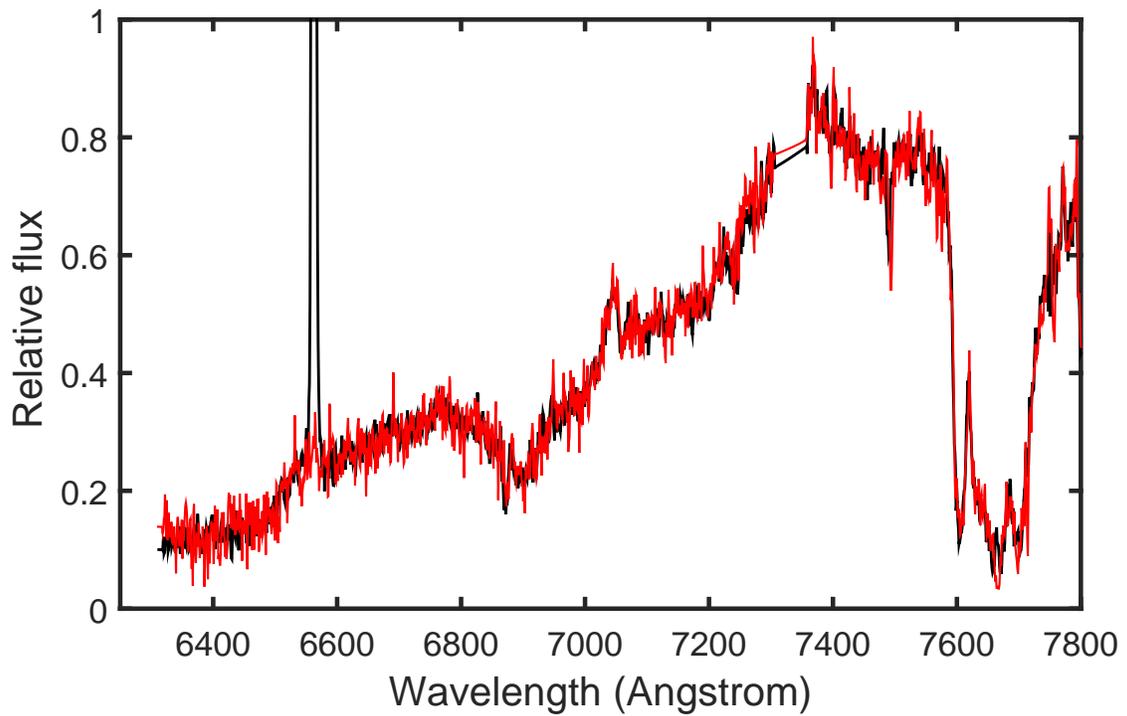}
\caption{Average of the two 2014 June 22 spectra of 2MASS~1520-4422 (black)
and of the two 2014 August 3 spectra (red). Note that the two continuum levels
are the same, despite the presence of the strong flare in the earlier spectra.}
\end{figure}

\end{document}